\documentclass[aps,prl,reprint,superscriptaddress,noofootinbib,showpacs,floatfix]{revtex4-1}

\usepackage{float,subfigure,graphicx,xcolor}

\usepackage{epstopdf}
\usepackage{amsmath}
\usepackage{amssymb}
\usepackage{placeins}
\usepackage{chemarrow}

\usepackage{cases}% http://ctan.org/pkg/cases

\usepackage{tikz}
\usepackage[T1]{fontenc}
\usetikzlibrary{matrix}

\begin{document}

\title{The heterodimer auto-repression loop: a robust and flexible
pulse-generating genetic module}

\author{B. Lannoo}
\affiliation{KU Leuven, Institute for Theoretical Physics, Celestijnenlaan
200D, 3001 Leuven, Belgium}
\affiliation{Laboratoire de Physique des Lasers, Atomes, et Mol\'ecules, 
UFR de Physique, Universit\'e Lille 1, Villeneuve d'Ascq, France}
\author{E. Carlon}
\affiliation{KU Leuven, Institute for Theoretical Physics, Celestijnenlaan
200D, 3001 Leuven, Belgium}
\author{M. Lefranc}
\affiliation{Univ. Lille, CNRS, UMR 8523~-~PhLAM~-~Physique des
  Lasers, Atomes et Mol\'ecules, F-59000 Lille, France}

\begin{abstract}
We investigate the dynamics of the heterodimer autorepression loop (HAL),
a small genetic module in which a protein $A$ acts as an auto-repressor
and binds to a second protein $B$ to form a $AB$ dimer. For suitable
values of the rate constants the HAL produces pulses of $A$ alternating
with pulses of $B$.  By means of analytical and numerical calculations,
we show that the duration of $A$-pulses is extremely robust against
variation of the rate constants while the duration of the $B$-pulses
can be flexibly adjusted.  The HAL is thus a minimal genetic module
generating robust pulses with tunable duration an interesting property
for cellular signalling.
\end{abstract}

\pacs{87.17.Aa, 87.16.Yc, 82.40.Bj, 87.18.Vf}

\date{\today}

\maketitle

Living cells regulate their response to stimuli through biochemical
reaction networks where genes, messenger RNAs (mRNAs) and proteins
interact with each other \cite{albe02}. Genes control the synthesis of
proteins via mRNAs, while their activities are regulated by specific
DNA-binding proteins called transcription factors (TF). Proteins bind
to each other to regulate their properties. These multiple interactions
are organized in entangled feedback loops, which generate a complex
and collective dynamics. Despite the high complexity of biological
networks, many specific dynamical mechanisms can be attributed to
small genetic modules comprising a few genes, their mRNAs and proteins
\cite{tyso03,bookAlon}. Thus, many studies have aimed to uncover the
dynamical design principles of such modules, viewed as building blocks
for larger systems or as devices for synthetic biology. For example,
the appearance of oscillations has been linked to negative feedback
and time delays \cite{tyso03}, and the importance of mechanisms
such as complexation \cite{fran05} or saturated degradation
\cite{kris06,degrade_and_fire,meng10} for oscillations has been
highlighted.

While much effort has been devoted to assess the robustness of biochemical
oscillations, it has generally been quantified only by the constancy of
the total period. The latter is an important criterion for oscillations
whose purpose is time keeping, as in circadian clocks, but it is not
always relevant. Recent studies (see \cite{purv13} for a review) revealed
that also signaling proteins, which detect and deliver cellular signals,
can display oscillating dynamics. In some systems, oscillations appear as
discrete pulses separated by constant time intervals \cite{laha08}, while
in others the intensity of upstream signals determines the time interval
between pulses \cite{hao12,Locke11:_stoch_pulse_regulation}, which may
thus be used to encode information~\cite{meng10}. A natural question is
then whether we can identify simple model systems that display similar
behavior. In this Letter, we investigate the dynamical properties of such
a minimal genetic module, the Heterodimer Autorepression Loop (HAL).
The HAL generates a periodic ``pulsating" output in the concentrations
of two different proteins where the pulses of one protein alternate with
the pulses of the other one. We will use the term ``pulses" rather than
``oscillations" to emphasize that we think primarily to the model as a
genetic device for cellular signalling, rather than for time keeping.
Remarkably, the duration of the pulses of one protein is robust against
variations in the rate constants, while the time interval between two
pulses, where the other protein is dominant, is tunable.

\begin{figure}[t]
\includegraphics[width=0.75\columnwidth]{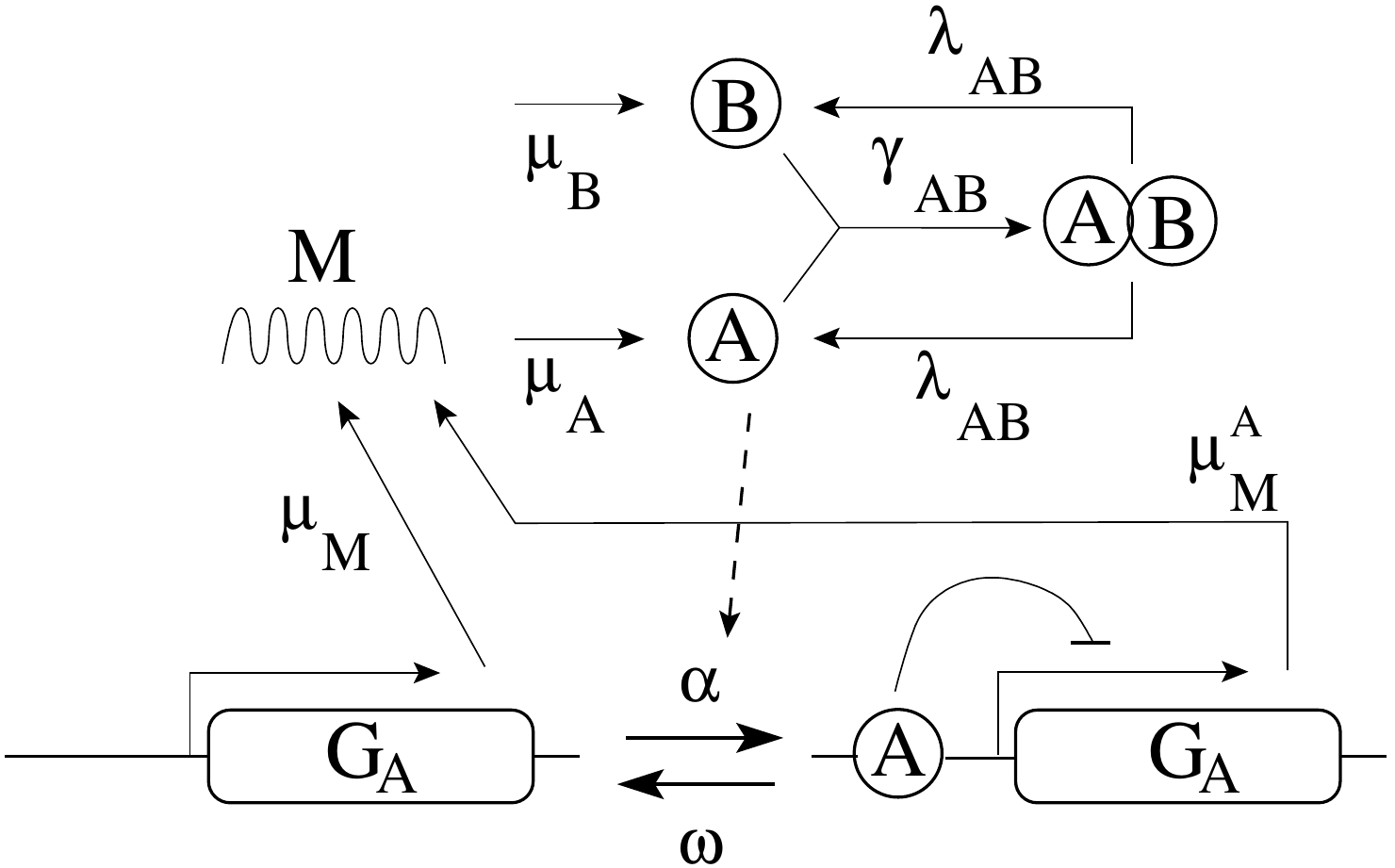}
\caption{A schematical representation of the HAL module. The gene $G_A$
is repressed by its own protein $A$, which forms an inactive dimer $AB$
with a second protein/molecule $B$. Proteins and mRNA degrade with
rates $\delta_A$, $\delta_B$, $\delta_{AB}$ and $\delta_M$ (reactions
not shown).}
\label{HAL_scheme}
\end{figure}

The HAL consists of a self-repressing TF protein $A$ that can bind to
its own gene to inhibit mRNA synthesis, or to another protein $B$,
then becoming inactive (Fig.~\ref{HAL_scheme}). Self-repression
is a pervasive motif in transcriptional
networks~\cite{Hermsen2010a,Salgado01012001,Keseler01012005},
and protein-protein interactions modifying TF activity are also
ubiquitous~\cite{Szklarczyk02112010}, making the HAL very plausible
biologically. Accordingly, the HAL appeared with high frequency in
evolutionary algorithm calculations searching for oscillating modules
\cite{dorp13}. The HAL can be described by the following deterministic
differential equations, obtained from the reactions in Supplemental
Fig.~1 using mass action kinetics:
\begin{eqnarray}
\left\{ 
\begin{array}{ccl}
\vspace{2mm}
\frac{d[G]}{dt}   &=&\omega (1-[G])   -\alpha [G] [A]        \\
\vspace{2mm}
\frac{d[M]}{dt}   &=&\mu_M [G]         + \mu_M^A(1-[G])       -\delta_M [M]          \\
\vspace{2mm}
\frac{d[A]}{dt}   &=&\mu_A [M]        -\delta_A [A]          -\gamma_{A\! B} [A] [B] 
\\ \vspace{2mm}
&& +\lambda_{A\! B} [AB]  +\omega (1-[G]) -\alpha [G] [A]\\
\vspace{2mm}
\frac{d[B]}{dt}   &=&\mu_B            -\delta_B [B]          -\gamma_{A\! B} [A] [B]
\vspace{2mm}
+\lambda_{A\! B} [AB]\\
\vspace{2mm}
\frac{d[AB]}{dt}  &=&\gamma_{A\! B} [A] [B] -\lambda_{A\! B} [AB] -\delta_{A\! B} [AB]
\end{array}
\right.
\label{full_model}
\end{eqnarray}
where $[A]$, $[B]$, $[AB]$ and $[M]$ are the concentrations of $A$, $B$,
$AB$ and of the mRNA produced by the gene $G_A$, respectively (since $G_B$
is unregulated, the concentration of its mRNA is not a variable). The
first equation in \eqref{full_model} describes the dynamics of gene $G_A$
activity, which is a continuous variable $0 \leq [G] \leq 1$, with $[G]=0$
(resp., $[G]=1$) when the gene is permanently protein-bound and repressed
(resp., unbound and active)~\cite{fran05,mora09}. Such an average activity
appears naturally in rate equations derived from a moment expansion of the
chemical master equation~\cite{Wang14:_stoch_oscil_selfrepressed_gene}. It
takes into account that due to transcriptional
bursting~\cite{Golding05,xavier07,Chubb06,Suter11:_mammalian_bursting,Harper11:_stochastic_cycles},
gene activity is out of equilibrium and lags variations in
TF concentration. The equation used here is valid only when
the gene response is not too slow compared to mRNA and protein
lifetimes~\cite{Wang14:_stoch_oscil_selfrepressed_gene}, thus the
predictions of our deterministic approach will be carefully checked with
stochastic simulations of the HAL.

%%%%%%%%%%%%%%%%%%%%%%%%%%%%%%%%%%%%%%%%%%%%%%%%%%%%%%%%%%%%%%%%
%%%%%%%%%%%%%%%%%%%%%%%%%%%%%%%%%%%%%%%%%%%%%%%%%%%%%%%%%%%%%%%%

%%%%%%%%%%%%%%%%%%%%%%%%%%%%%%%%%%%%%%%%%%%%%%%%%%%%%%%%%%%%%%%%
\begin{table}
\caption{\label{tab:table1} 
Typical biological ranges for rate constants in the model, as
obtained from the literature. The last three parameters are guessed.
$\delta$'s and $\mu$'s are the degradation and synthesis rates,
respectively. $\gamma_{A\! B}$ and $\lambda_{A\! B}$ are the association
and dissociation constants of the AB dimer.  $\alpha$ and $\omega$ are
the binding and unbinding rates of the protein A to the gene.  The ratio
$[A]_0\equiv \omega/\alpha$ defines a regulation threshold: for $[A]
\gg [A]_0$ the promotor region has a protein bound to it, while for
$[A] \ll [A]_0$ the promotor is free.  The system is considered to be
enclosed in a cell of volume $V=50\mu m^3$.  We take this as as volume
unit. The concentration $[X]$ of a species $X$ then correspond to the
number of molecules $X$ in $V$.  All values are expressed in minutes,
except for $[A]_0$ which is a dimensionless number.}
\begin{ruledtabular}
\begin{tabular}{ccc}
Parameter & Value & Reference \\
\hline
$1/\mu_M$ & $[0.1,100]$ & \cite{mRNA_production}\\
$1/\delta_M$ & $[3, 60]$ & \cite{mRNA_degradation}\\
$1/\mu_A$ & $[10^{-4},10]$ & \cite{protein_production}\\
$1/\mu_B$ & $[10^{-3},100]$\footnote{Obtained from the value of
$\mu_A$, and assuming a typical number of 10 mRNA's in the cell.} &
\cite{protein_production}\\
$1/\delta_A$, $1/\delta_B$, $1/\delta_{A\! B}$ & $[4, 2000]$ &
\cite{protein_degradation}\\
$1/\omega$ & $[1,60]$ & \cite{Poorey2013}\\
$1/\gamma_{A\! B}$ & $[0.02, 20]$ & \cite{fran05}\footnote{Assuming that
the formation of the AB complex is diffusion limited and $D= 1 {\rm \mu
m^2 \cdot s^{-1}}$.}\\
$1/\lambda_{A\! B}$ & $100$ & - \footnote{This choice implies a small
dissociation rate, so that the complex is irreversibly formed.}\\
$1/\mu_M^A$ & $10^3$ & - \footnote{This is the transcription rate from
a gene with the protein a bound. For an ideal repressor $\mu_M^A=0$,
we assume here that there is a weak transcription even with the protein
bound. This rate is however at least $10$ smaller that the transcription
rate from a free gene (see value of $1/\mu_M$ above).} \\
$[A]_0$ & $[1,100]$ & -
\footnote{Here it is assumed that one needs from 1 to 100 proteins in
the volume at threshold to bind to the gene.}\\
\end{tabular}
\end{ruledtabular}
\end{table}

%%%%%%%%%%%%%%%%%%%%%%%%%%%%%%%%%%%%%%%%%%%%%%%%%%%%%%%%%%%%%%%%%%%%%%%%%% 

To explore the dynamics of the HAL, the rate constant values were randomly
sampled in typical biological ranges obtained from the litterature
\cite{mRNA_production,mRNA_degradation,protein_degradation,protein_production,Poorey2013},
as shown in Table~\ref{tab:table1}.  Robust pulses were
found in a significant domain of parameter space (Supplemental
Figure~2).  As a general rule, pulses are observed if
$\gamma_{A\! B}$ is large while $\lambda_{A\!  B}$ is small, so that
the complex is irreversibly formed (large or small meaning close to
the upper or lower bound in Table~\ref{tab:table1}). Also, the protein
production rates $\mu_A$ and $\mu_B$ need to be sufficiently large and
to verify $\mu_B \lesssim \mu_A \mu_M/\delta_M$. The latter condition
expresses that the productions of $A$ and $B$ should be balanced, with
$A$ synthesized faster than $B$ for a fully active gene ($[G]=1$, with
mRNA concentration $[M]=\mu_M/\delta_M$), and more slowly for an inactive
gene. The average period was $T_{\rm tot} \approx 100\, \text{min}$.

\begin{figure}[t]
\includegraphics[width=\columnwidth]{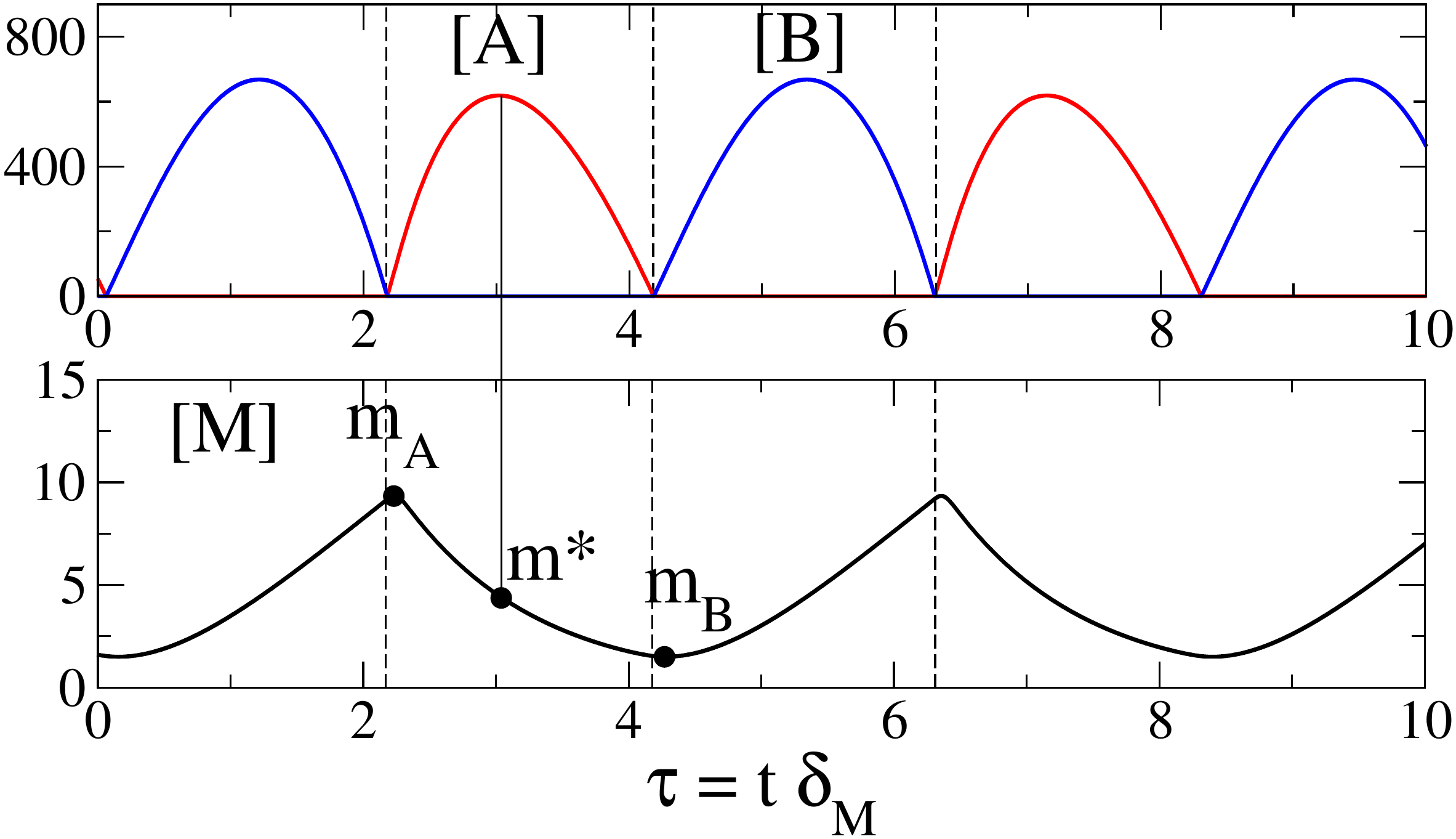}
\caption{Protein (top) and mRNA (bottom) concentrations vs.
time (in units of the characteristic mRNA degradation time
$\delta_M^{-1}$) for the following parameter values: $\mu_M^{-1} =
0.5$, $\delta_M^{-1} = 20$, $\mu_A^{-1} = 0.067$, $\mu_B^{-1} =
0.015$, $\delta_A^{-1} = \delta_B^{-1} = 10^3$, $\delta_{A\!
B}^{-1} = 10$, $\gamma_{A\! B}^{-1} = 0.02$, $\omega^{-1} = 100$,
$[A]_0 = 1$, ($\lambda_{A\! B}$ and ${\mu_M^A}$ are fixed as in
Table~\ref{tab:table1}). A (resp., B) concentration is plotted in
red (resp., blue). Dashed lines indicate the beginning of the A- and
B-phases. During the A-phase the mRNA concentration decays as the
$A$ protein strongly represses its own gene.}
\label{exampleHAL}
\end{figure}

Figure~\ref{exampleHAL} shows a typical pulsating solution of
\eqref{full_model}, with a total period $T_{\rm tot} = 64\, \text{min}$.
The mutual ``sequestration'' of $A$ and $B$ induced by the dimerization
leads to an alternation of pulses where either $A$ or $B$ is predominant
(referred to as the $A$- and $B$-phase), the other protein remaining at
low levels. Inside each pulse, the dominant protein first accumulates
as it is synthesized faster than the other while complexation removes
the two proteins in equal quantities. Then, it decreases to almost
zero when the situation is reversed. During the $B$-phase, the gene is
unrepressed, and $A$ synthesis rate increases as mRNA builds up. During
the $A$-phase, the gene is repressed and $A$ synthesis rate decreases as
mRNA is degraded. The key for cycling is thus that during each phase,
there is a time where $A$ and $B$ synthesis rates become equal, which
is at the peak of the pulse.

Thus, mRNA life time plays the role of a time delay, a crucial ingredient
for oscillations \cite{nova08}. The sequestration of the TF A also plays
an important role by inducing an ultrasensitive response in gene activity
\cite{buch09}, a strong nonlinear effect~\cite{GoldbeterKoshland81}
which favors oscillations like a high transcriptional cooperativity
would do. This ultrasensitivity is presumably also important in other
gene circuits where sequestration induces oscillations \cite{fran05}.

%%%%%%%%%%%%%%%%%%%%%%%%%%%%%%%%%%%%%%%%%%%%%%%%%%%%%%%%%%%%%%%%%%%%%%
%%%%%%%%%%%%%%%%%%%%%%%%%%%%%%%%%%%%%%%%%%%%%%%%%%%%%%%%%%%%%%%%%%%%%%

To get an estimate of the pulses period, we make some
simplifications. We assume  perfect repression ($\mu_M^A=0$) and 
irreversible complex formation ($\lambda_{A\!  B}=0$). With the
latter assumption, we do not need to track dimer $AB$, leading from
Eqs.~\eqref{full_model} to a system of four differential equations
only.  Considering that proteins dimerize before they degrade, we
set $\delta_A=\delta_B=0$. We neglect the variation of $[A]$ due to
the binding or unbinding of one molecule, which removes the terms
involving $[G]$ in the equation for $d[A]/dt$ in (\ref{full_model}).
Rescaling the time as $\tau\equiv t \delta_M$ and the concentrations
as $a \equiv [A] \gamma_{A\! B}/\delta_M$, $b \equiv [B] \gamma_{A\!
B}/\delta_M$, $m \equiv [m] \delta_M/\mu_M$ and $g=[G]$, one gets:
\begin{eqnarray}
\left\{ 
\begin{array}{ccl}
\frac{dg}{d\tau}\,   &=& \Omega (1-g)       -\sigma g a \\
\frac{dm}{d\tau}     &=& \quad g     \qquad -       m   \\
\frac{da}{d\tau}\,   &=& \quad k_a  m \quad -       a b \\
\frac{db}{d\tau}\,   &=& \quad k_b   \qquad -       a b 
\end{array}
\right.
\label{HAL_reduced}
\end{eqnarray}
where the rescaled parameters are $\Omega \equiv
\omega/\delta_M$, $\sigma \equiv \alpha/\gamma_{A\! B}$, $k_a \equiv
\mu_A\mu_M\gamma_{A\! B}/\delta_M^3$ and $k_b \equiv \mu_B\gamma_{A\!
B}/\delta_M^2$. There is no protein degradation in \eqref{HAL_reduced},
but the irreversible complexation $A+B \to AB$ prevents unbounded growth.

Assuming total repression in the $A$-phase ($g=0$) and slow unbinding
of $A$ from the gene in the $B$-phase (small $\Omega$) we get the
following two equations for $T_a$ and $T_b$, the durations of the $A$-
and $B$-phase, respectively (Supplemental Material):
\begin{subequations}
\begin{eqnarray}
\frac{T_a}{e^{T_a}-1} &=&  \beta \frac{-1+T_b+e^{-T_b}}{e^{T_a}-e^{-T_b}}
\label{per1_txt}\\
\frac{T_a}{e^{T_a}-1} &=& 
\frac{\beta \left(T_b - \frac{T_b^2}{2}\right) +  T_b}{1 - e^{-T_b}} - \beta
\label{per2_txt}
\end{eqnarray}
\label{per12_txt}
\end{subequations}
which depend on a single parameter
\begin{equation}
\beta \equiv \frac{k_a \Omega}{k_b} = 
\frac{\omega}{\delta_M} \frac{(\mu_A\mu_M)/\delta_M}{\mu_B}
\label{beta}
\end{equation}
which is the ratio of mRNA lifetime to gene response time,
multiplied by the ratio of maximal A synthesis rate to B synthesis
rate. 

%%%%%%%%%%%%%%%%%%%%%%%%%%%%%%%%%%%%%%%%%%%%%%%%%%%%%%%%%%%%%%%%%%%%%%%%%
\begin{figure}[t]
\includegraphics[width=\columnwidth]{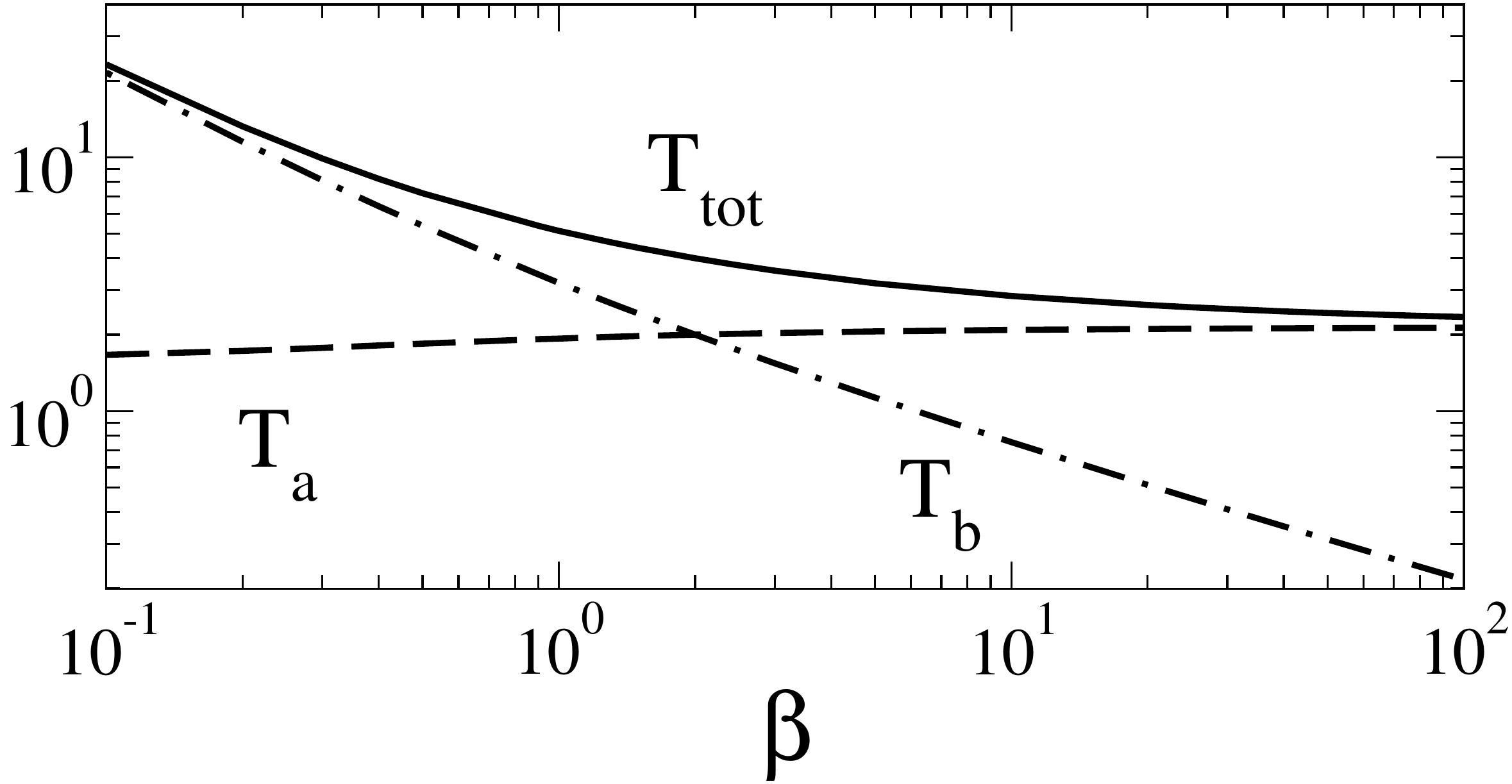}
\caption{Log-log plot of $T_a$ (dashed line) and $T_b$ (dot-dashed line),
the solutions of Eqs.~(\ref{per12_txt}) as a function of $\beta$. The
total period $T_{\rm tot} = T_a+T_b$  is shown as a solid line.}
\label{final_fit}
\end{figure}
%%%%%%%%%%%%%%%%%%%%%%%%%%%%%%%%%%%%%%%%%%%%%%%%%%%%%%%%%%%%%%%%%%%%%%%%

Figure~\ref{final_fit} displays $T_a$, $T_b$ and the total
period $T_{\rm tot} = T_a+T_b$, obtained by numerically solving
Eqs.~(\ref{per12_txt}). Remarkably, $T_a$ depends little on $\beta$,
varying by about $30\%$ ($1.67 \leq T_a \leq 2.13$) when $\beta$ changes
over three orders of magnitude ($10^{-1} \leq \beta \leq 10^2$). On
the contrary, $T_b$ is very sensitive to $\beta$ and ranges over two
orders of magnitude.  The pulses of $A$ are ``robust'', i.e. of almost
constant duration, while the duration of the $B$-pulses can be tuned
by changing $\beta$. Hence, any parameter which $\beta$ depends on
(see Eq.~\eqref{beta}) can be used to regulate the separation between
the pulses of $A$.

A detailed analysis of Eqs.~\eqref{per12_txt} is presented in the
Supplemental Material. Here we give simple arguments
explaining the main features observed. During the $A$-phase, $m(\tau)$
decays exponentially [set $g=0$ in Eqs.~\eqref{HAL_reduced}]. Denoting
by $m_A$ and $m_B$ the mRNA concentrations at the beginnings of the
$A$- and $B$-phases (Fig.~\ref{exampleHAL}), we have $m_B=m_A
e^{-T_a}$. To get pulses, $A$ synthesis must be faster than
$B$ synthesis when $A$-phase starts ($k_a m_A > k_b$), and slower when
B-phase starts ($k_a m_B < k_b$), which yields $m_B<k_b/k_a<m_A$.
Assuming stationarity of the $B$ protein ($db/d\tau\sim0$) in the
$A$-phase, we get
\begin{equation}
  \label{eq:dadtau}
\frac{da}{d\tau} = k_a m(\tau) - k_b = k_a m_A e^{-\tau} - k_b 
\end{equation} 
The solution of \eqref{eq:dadtau} is a pulse with a peak (${da}/{d\tau}
=0$) at mRNA concentration $m^*=k_b/k_a$ (Fig.~\ref{exampleHAL}). The
pulse duration $T_a$ is found by setting $a(T_a)=0$:
\begin{equation} 
\frac{T_a}{1-e^{-T_a}} = \frac{k_a m_A}{k_b}
\label{Ta_pap}
\end{equation}
Hence, $T_a$ depends only on the ratio $k_a m_A/k_b$. Since pulses require
$k_a m_A/k_b > 1$, $T_a$ cannot become too small.  Eq.~\eqref{Ta_pap}
might suggest that large values of $k_a/k_b$ lead to arbitrarily
large $T_a$. However, this is not true because the $B$-phase shrinks
as $k_a/k_b$ gets larger, since $B$ synthesis is then faster than $A$
synthesis only for a short time. Hence the variations of $m$ during
the $B$-phase become smaller and smaller as $k_a/k_b$ increases,
since the mRNA characteristic time is $1$.  Consequently, $m_B/ m_A =
e^{-{T_a}}$ remains close to $1$, thus bounding $T_a$. In simple words,
changes in the rate constants which could affect $T_a$ are compensated
by a associated change in the mRNA maximum concentration $m_A$. Thus,
there is a natural negative feedback loop stabilizing $A$-pulse duration.

%%%%%%%%%%%%%%%%%%%%%%%%%%%%%%%%%%%%%%%%%%%%%%%%%%%%%%%%%%%%%%%%%%%%%%%%%%%%
\begin{figure}[t!]
\includegraphics[width=0.95\columnwidth]{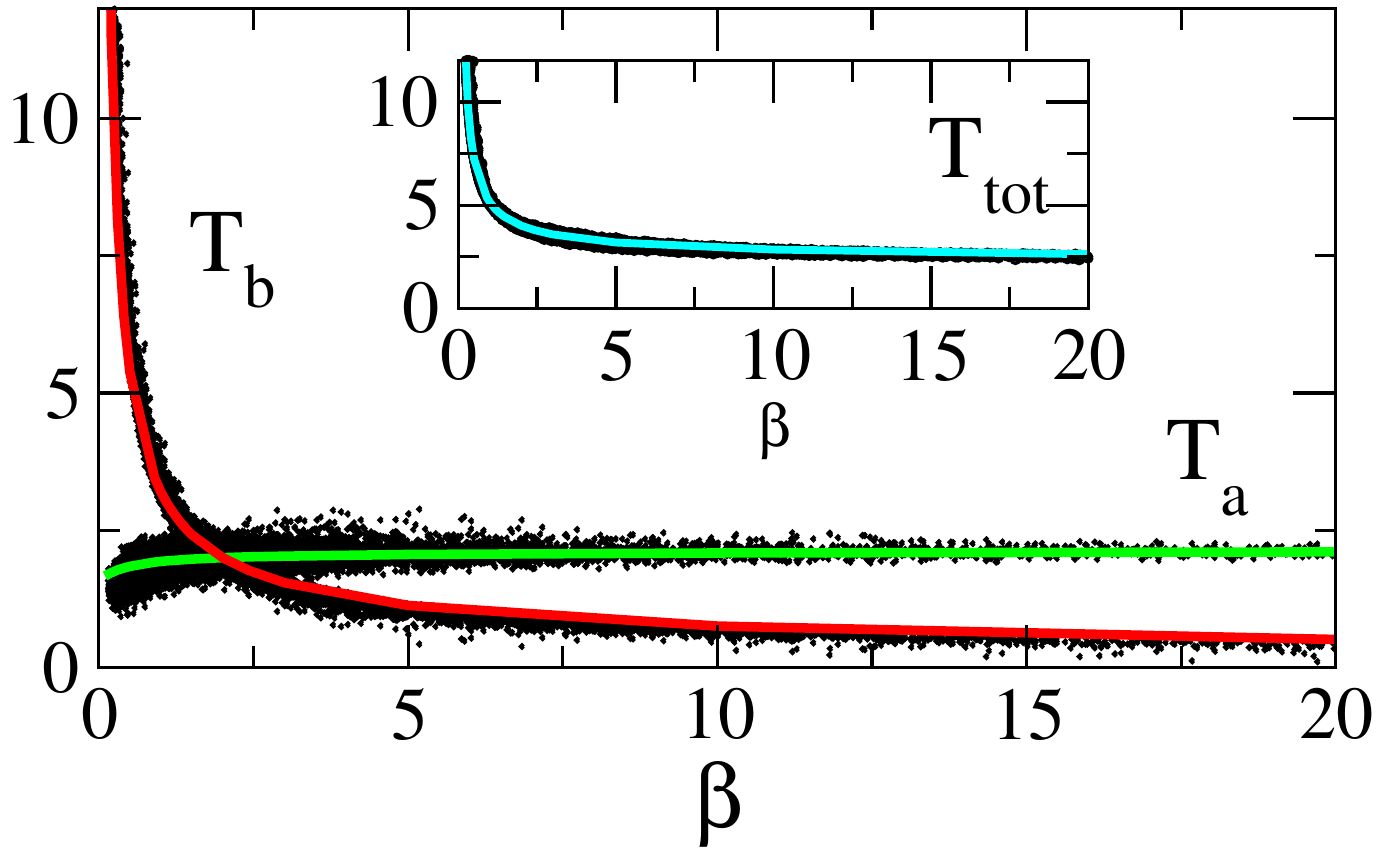}
\caption{Solid lines: analytical estimates of $T_a$ and $T_b$
from Eqs.~(\ref{per12_txt}). Circles: durations of the A and B phases as
computed from the numerical integration of \eqref{full_model}.
Inset: Comparison for the total period $T_\text{tot}=T_a+T_b$.}
\label{more_numerics}
\end{figure}
%%%%%%%%%%%%%%%%%%%%%%%%%%%%%%%%%%%%%%%%%%%%%%%%%%%%%%%%%%%%%%%%%%%%%%%%%%%

To corroborate these results, based on the reduced model
\eqref{HAL_reduced} and further approximations, we numerically
computed $T_a$ and $T_b$ using the full equations~\eqref{full_model}
for parameter sets $\{k_i\}$ centered around the set $\{k_i^0\}$ used in
Fig.~\ref{exampleHAL}. Each $k_i$ was selected randomly and uniformly on
a logarithmic scale in the interval $[\frac 12 k_i^0,2k_i^0]$. In total
$10^3$ sets were generated, of which $98\%$ had a pulsating output,
showing that the parameter set of Fig.~\ref{exampleHAL} is well inside
the pulsating domain in parameter space. Although the data span a wide
range of values of $\beta$, the computed values of $T_a$, $T_b$ and
$T_{\rm tot}$ are in close agreement with the analytical approximation
(Fig.~\ref{more_numerics}).

\begin{figure}[t]
\includegraphics[width=\columnwidth]{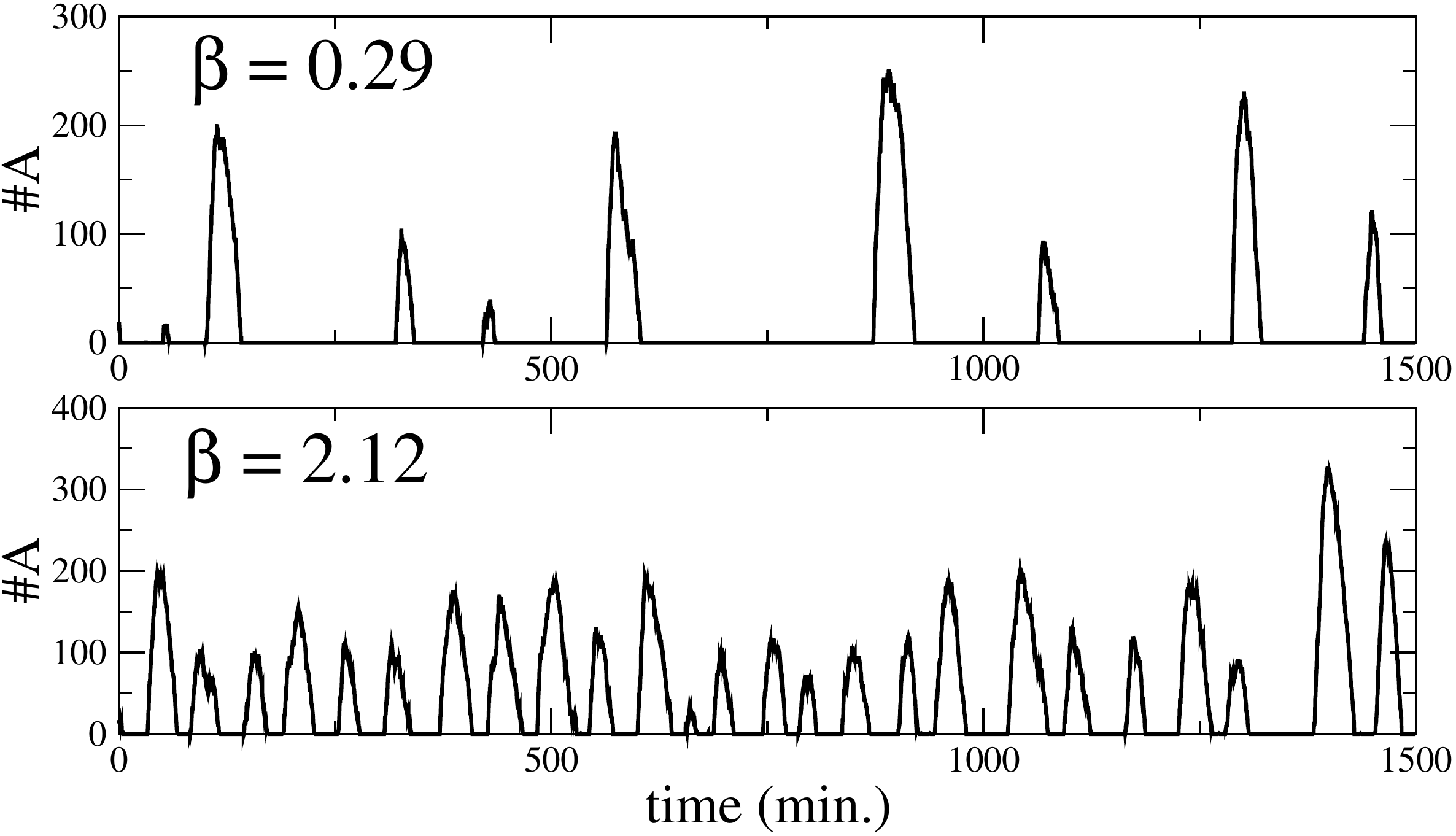}
\caption{
Stochastic simulations of the HAL for low and high values of
$\beta$, corresponding to short and long time intervals between
A-pulses, respectively. Parameters of the top graph:
$\mu_M^{-1} = 1.11$, $\delta_M^{-1} = 16.67$, $\mu_A^{-1} = 0.59$, 
$\mu_B^{-1} = 0.05$, $\delta_A^{-1} = \delta_B^{-1} = 10^3$,
$\delta_{A\!B}^{-1} = 10$, $\gamma_{A\! B}^{-1} = 0.02$,
$\omega^{-1} = 73.11$, $[A]_0 = 1$.
For the bottom graph the parameters are the same except
$\omega^{-1} = 10$.
}
\label{stochsim}
\end{figure}

A legitimate question is then whether our findings still hold true
when the stochastic nature of biochemical networks cannot be ignored,
especially since a slow promoter dynamics may be needed to obtain
long intervals between A-pulses. We therefore carried out stochastic
simulations of the reaction network of Fig.~\ref{HAL_scheme}, using the
Gillespie algorithm~\cite{gill77}.  Pulses are observed for both high
and low values of $\beta$, with a stable time interval between A-pulses
(Fig.~\ref{stochsim} and Supplemental Material), which confirms the
relevance of our analysis.

Summarizing, we have investigated the dynamics of the HAL, a
pulse generator based on the competing effects of self-repression
and complexation. Self-repression alone does not typically induce
oscillations, unless time delays \cite{stri08} or strong nonlinearities
are introduced. Protein complexation generates an effective ultrasensitive
response \cite{buch09} which can induce oscillations as in other examples
\cite{fran04}, including the mixed-feedback loop \cite{fran05} or the
monomer-dimer oscillator \cite{dorp13}. Since the only role of $B$ is to
sequester A, B does not need to be a protein but could be any inhibitor
molecule binding to A to block its transcriptional activity.

A striking feature of the HAL is that the duration of the the
A-pulses is robust against variation of the rate constants, whereas
the duration of the B-pulses is tunable.  It has been suggested that
biological signals may be encoded in time interval between pulses
\cite{meng10,laha08,hao12,purv13,Locke11:_stoch_pulse_regulation}.
Since the HAL is a robust and flexible pulse generator, it would perfectly
fit into this design.

The self-repression motif is highly represented in genetic networks
\cite{bookAlon}. It would be interesting to see if the HAL, a
simple extension of this motif, is also ubiquitous. Known examples
of oscillations based on a self-repressing protein $A$ have been
attributed to delay or high cooperativity, perhaps sometimes obscuring
the implication of a binding partner $B$.  A closely related oscillator
is the Mixed-feedback loop (MFL) \cite{fran05}, which is also based on
a $AB$ dimer formation, but the protein $A$ activates the transcription
of gene $G_B$ instead of repressing itself. Interestingly, an analysis
of E.  coli motifs involving both transcriptional and protein-protein
interactions led to the discovery of the MFL but since it excluded
self-repression, was not able to detect the HAL~\cite{yege03}. The
MFL network motif is overrepresented in Yeast cells \cite{yege03}
and is also at the core of circadian clocks in Mammals, Neurospora or
Drosophila \cite{fran05}. It is natural to expect that the HAL, being
closely related to the MFL, is also the core component of some natural
biochemical oscillators. Its simplicity, and interesting dynamical
properties also make the HAL a promising module for synthetic biology.

%%%%%%%%%%%%%%%%%%%%%%%%%%%%%%%%%%%%%%%%%%%%%%%%%%%%%%%%%%%%%%%%%%%%%%%%%%%%
%%%%%%%%%%%%%%%%%%%%%%%%%%%%%%%%%%%%%%%%%%%%%%%%%%%%%%%%%%%%%%%%%%%%%%%%%%%

\begin{acknowledgments}
We thank O. Biham, M. van Dorp, M. Nitzan, Q. Thommen, and B. Pfeuty
for discussions. Financial support from KU Leuven Grant No. OT/11/063 is
gratefully acknowledged, as well as by French Ministry of Higher Education
and Research, Nord-Pas de Calais Regional Council and FEDER through the
Contrat de Projets \'Etat-R\'egion (CPER) 2007--2013, and by LABEX CEMPI
(ANR-11-LABX-0007) operated by ANR.
\end{acknowledgments}

%  \bibliography{my_bib}


\vfill\eject

\section*{Supplemental Material}

In this document we provide a detailed analysis of various properties
of the HAL module.

\section{Analysis of full model}

We first consider the full model, which is given by:
\begin{eqnarray}
\left\{ 
\begin{array}{ccl}
\vspace{2mm}
\frac{d[G]}{dt}   &=&\omega (1-[G])   -\alpha [G] [A]        \\
\vspace{2mm}
\frac{d[M]}{dt}   &=&\mu_M [G]         + \mu_M^A(1-[G])       -\delta_M [M]          \\
\vspace{2mm}
\frac{d[A]}{dt}   &=&\mu_A [M]        -\delta_A [A]          -\gamma_{A\! B} [A] [B] 
\\ \vspace{2mm}
&& +\lambda_{A\! B} [AB]  +\omega (1-[G]) -\alpha [G] [A]\\
\vspace{2mm}
\frac{d[B]}{dt}   &=&\mu_B            -\delta_B [B]          -\gamma_{A\! B} [A] [B]
\vspace{2mm}
+\lambda_{A\! B} [AB]\\
\vspace{2mm}
\frac{d[AB]}{dt}  &=&\gamma_{A\! B} [A] [B] -\lambda_{A\! B} [AB] -\delta_{A\! B} [AB]
\end{array}
\right.
\label{suppl:full_model}
\end{eqnarray}
Table~\ref{BiochemicalReactions} lists all the reactions of the HAL
module shown in Fig. 1 of the main text and the corresponding mass action
terms. There are $12$ rate constants.

The first equation governs the time evolution of a variable $[G]$
which represents an average gene activity. Even in the cases where
gene activity is considered as a stochastic variable alternating
between two values (active and inactive), such equations
can be derived from moment expansions of the chemical master
equation~\cite{Wang14:_stoch_oscil_selfrepressed_gene}. The
form used here is valid when the variances of the
stochastic variables can be neglected. If the results of
~\cite{Wang14:_stoch_oscil_selfrepressed_gene} can be transposed here,
this would be the case when $\omega/\delta_M \geq 1$. However, we have
checked with stochastic simulations that this heuristic bound is too
pessimistic, because the main discrepancy observed for lower values of
$\omega/\delta_M$ is only a slightly higher variability in interpulse
time intervals. Thus, Equations~\eqref{suppl:full_model} are adequate
for most parameter sets considered in our analysis.

In the limit of fast gene dynamics, the quasi-steady-state 
approximation $d[G]/dt=0$ yields $[G]= (1+\alpha[A]/\omega)^{-1}$. Substituting this in
\eqref{suppl:full_model}, we get a system of four equations:
\begin{eqnarray}
\left\{ 
\begin{array}{ccl}
\vspace{2mm}
\frac{d[M]}{dt}   &=&\frac{\omega \mu_M + \alpha \mu_M^A [A]}{\omega + \alpha [A]}       
-\delta_M [M] \\
\vspace{2mm}
\frac{d[A]}{dt}   &=&\mu_A [M] -\delta_A [A] -\gamma_{A\! B} [A] [B] 
+\lambda_{A\! B} [AB]\\
\vspace{2mm}
\frac{d[B]}{dt}   &=&\mu_B            -\delta_B [B]          -\gamma_{A\! B} [A] [B]
+\lambda_{A\! B} [AB]\\
\vspace{2mm}
\frac{d[AB]}{dt}  &=&\gamma_{A\! B} [A] [B] -\lambda_{A\! B} [AB] -\delta_{A\! B} [AB]
\end{array}
\right.
\label{full_model_MM}
\end{eqnarray}
which recovers the standard Michaelis-Menten form for the mRNA synthesis.

%%%%%%%%%%%%%%%%%%%%%%%%%%%%%%%%%%%%%%%%%%%%%%%%%%%%%%%%%%%%%%%%%%%%%%%%%%%%%%%%%%%%%%%%%%%%%%%%%
\begin{figure}[t]
\begin{tikzpicture}
\clip node (m) [matrix,matrix of nodes,
fill=black!20,inner sep=0pt,
nodes in empty cells,
nodes={minimum height=0.6cm,minimum width=0 .5cm,anchor=center,outer sep=0,font=\sffamily},
row 1/.style={align=center,nodes={fill=black,text=white}},
column 1/.style={text width=2.7cm,align=left,every even row/.style={nodes={fill=white}}},
column 2/.style={text width=1.7cm,align=center,every even row/.style={nodes={fill=white}}},
column 3/.style={text width=0.9cm,align=center,every even row/.style={nodes={fill=white}}},
column 4/.style={text width=1.8cm,align=center,every even row/.style={nodes={fill=white}}},
column 5/.style={text width=1.7cm,align=center,every even row/.style={nodes={fill=white}}},
prefix after command={[rounded corners=4mm] (m.north east) rectangle (m.south west)}
] {
	Reaction        & Reactants     & $\xrightarrow{\mbox{rate}}$     & Products      & Term\\
	$\ G$        transcription: & $G$           & $\xrightarrow{\mu_M}$           & $G + M$       &
	$\mu_M [G]$\\
	$\ M$        degradation:   & $M$           & $\xrightarrow{\delta_M}$        & $\varnothing$ &
	$\delta_M [M]$\\
	$\ M$        translation:   & $M$           & $\xrightarrow{\mu_A}$           & $M + A$       &
	$\mu_A [M]$\\
	$\ A$        degradation:   & $A$           & $\xrightarrow{\delta_A}$        & $\varnothing$ &
	$\delta_A [A]$\\
	$\ B$        production:    & $\varnothing$ & $\xrightarrow{\mu_B}$           & $B$           &
	$\mu_B$ \\
	$\ B$        degradation:   & $B$           & $\xrightarrow{\delta_B}$        & $\varnothing$ &
	$\delta_B [B]$\\
	$\ G$        repression:    & $G + A$       & $\xrightarrow{\alpha}$          & $G_{\! A}$    &
	$\alpha [G][A]$\\
	$\ G_{\! A}$ deregulation:  & $G_{\! A}$    & $\xrightarrow{\omega}$          & $G + A$       &
	$\omega (1-[G])$\\
	$\ G_{\! A}$ transcription: & $G_{\! A}$    & $\xrightarrow{\mu_M^A}$         & $G_{\! A} + M$&
	$\mu_M^A (1-[G])$\\
	$\ AB$       complexation:  & $A + B$       & $\xrightarrow{\gamma_{A\! B}}$  & $AB$          &
	$\gamma_{A\! B}[A][B]$\\
	$\ AB$       dissociation:  & $AB$          & $\xrightarrow{\lambda_{A\! B}}$ & $A + B$       &
	$\lambda_{A\! B}[AB]$\\
	$\ AB$       degradation:   & $AB$          & $\xrightarrow{\delta_{A\! B}}$  & $\varnothing$ &
	$\delta_{A\! B}[AB]$\\
};
\end{tikzpicture}
\caption{List of all the biochemical reactions which define the HAL
module. By convention rates are denoted by: $\mu$ for production rates,
$\delta$ for degradation rates, $\alpha$ for binding rates, $\omega$
for unbinding rates, $\gamma$ for complexation rates and $\lambda$
for dissociation rates. The rightmost column gives the corresponding rates
in the differential equations as obtained from mass action kinetics.}
\label{BiochemicalReactions}
\end{figure}
%%%%%%%%%%%%%%%%%%%%%%%%%%%%%%%%%%%%%%%%%%%%%%%%%%%%%%%%%%%%%%%%%%%%%%%%%%

\begin{figure*}[t]
\includegraphics[width=0.45\columnwidth]{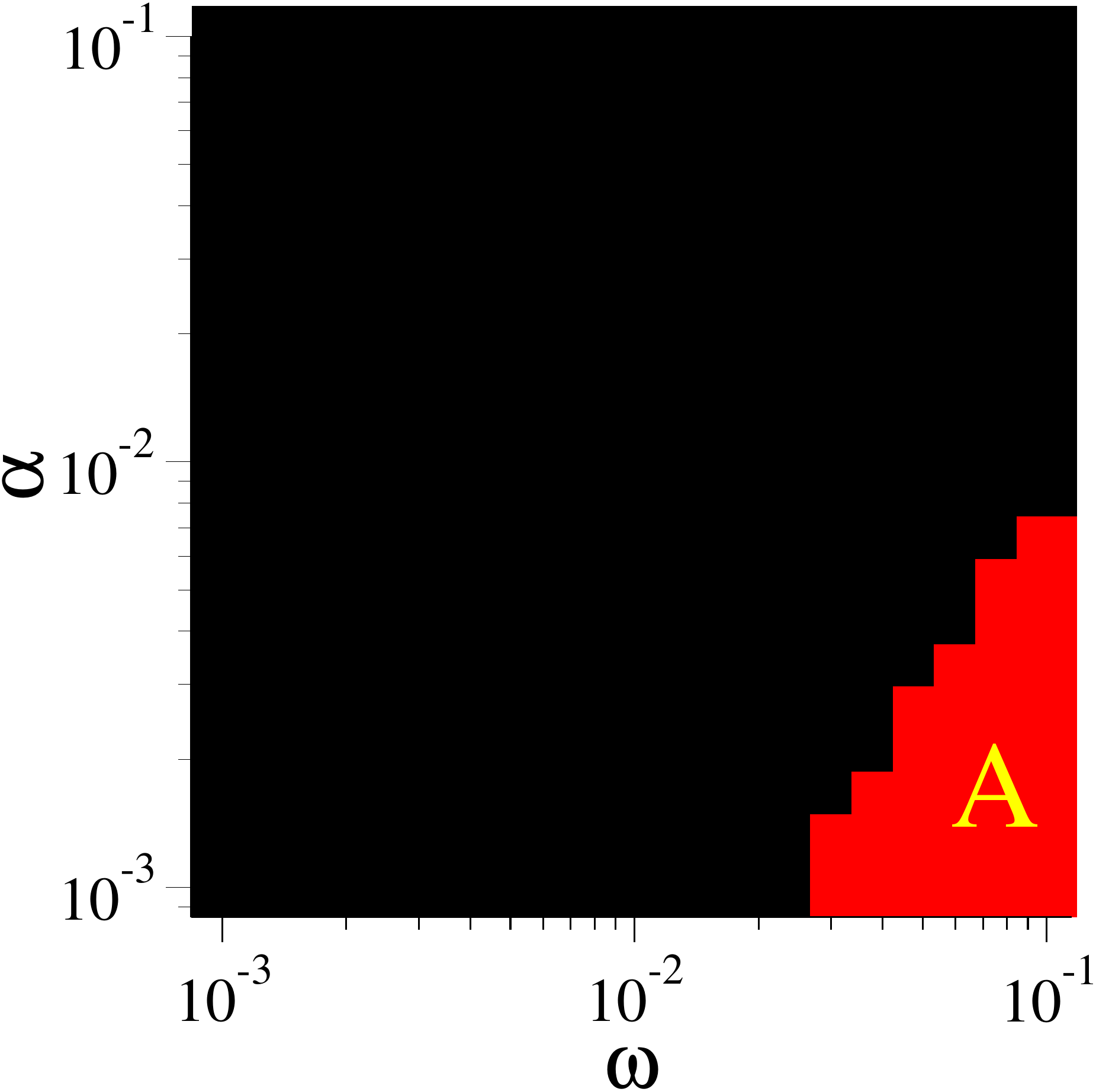}
\includegraphics[width=0.45\columnwidth]{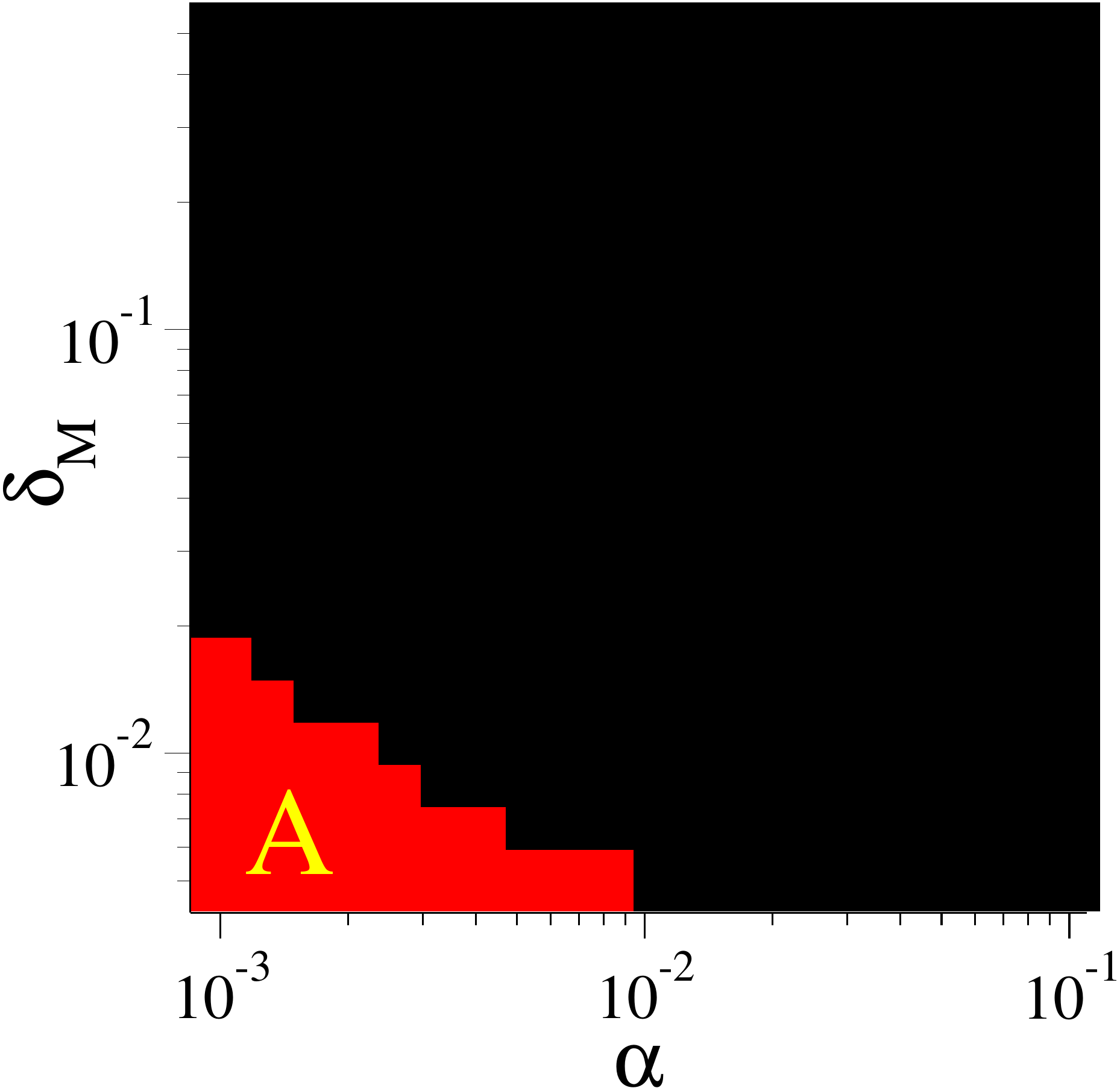}
\includegraphics[width=0.45\columnwidth]{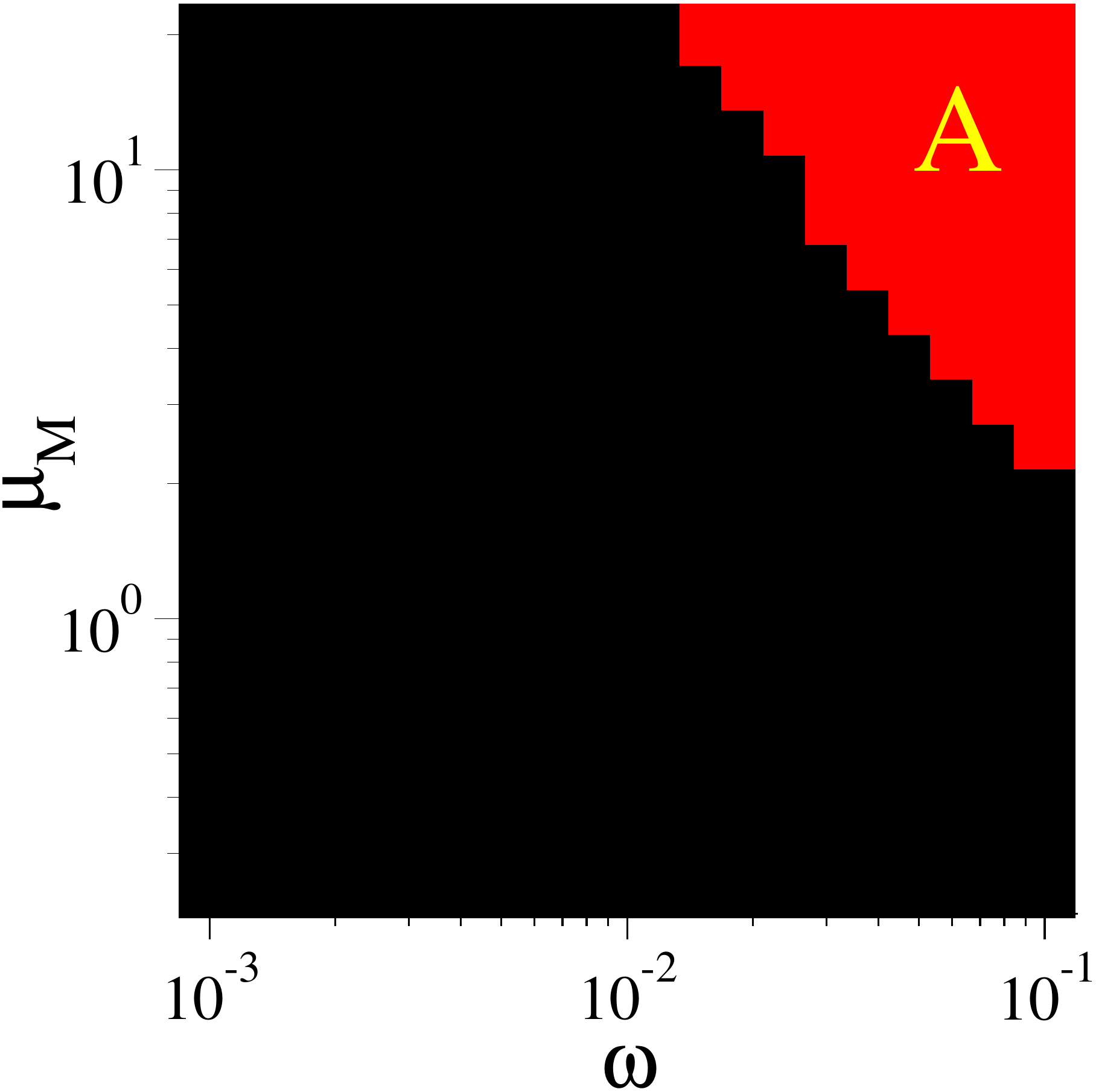}

\includegraphics[width=0.45\columnwidth]{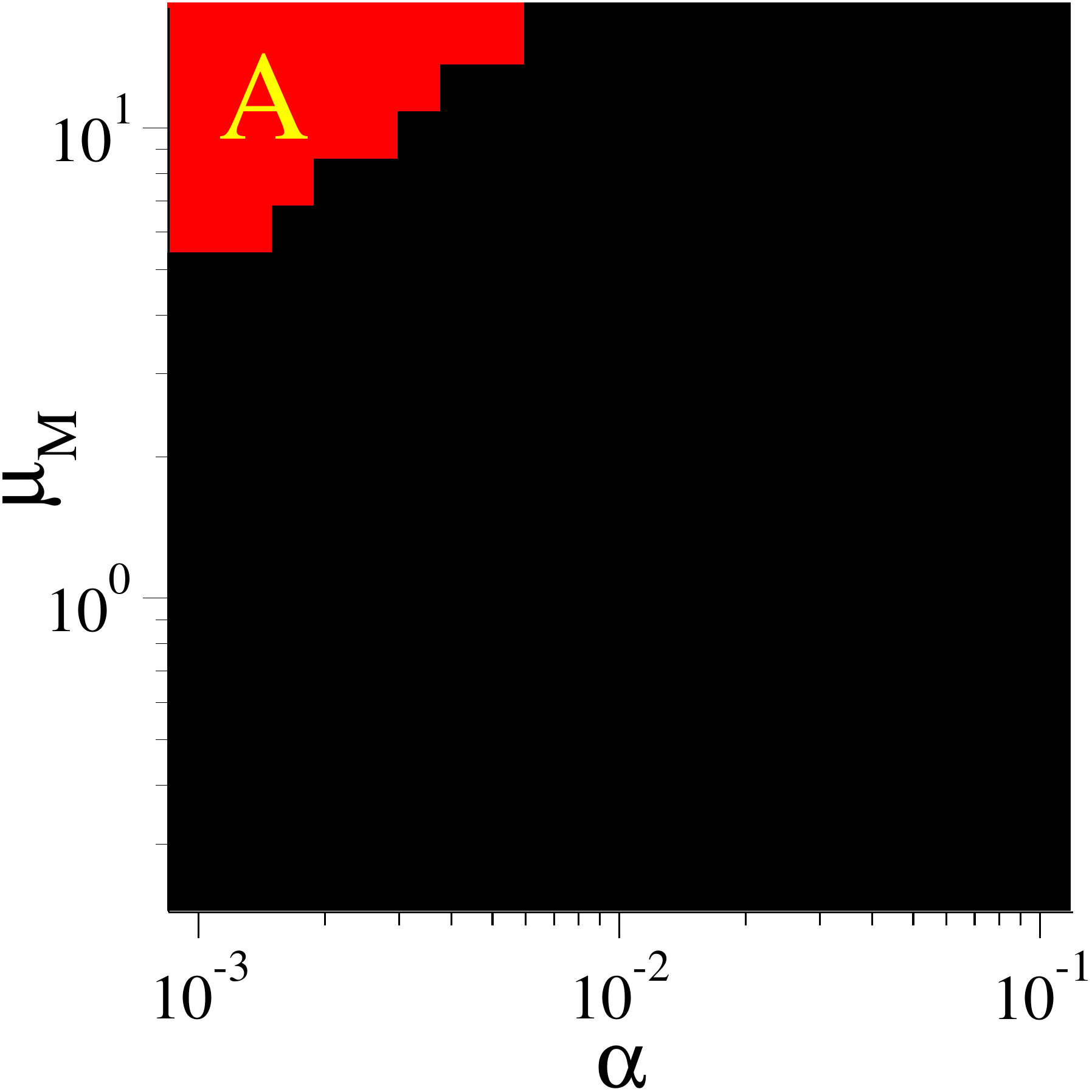}
\includegraphics[width=0.45\columnwidth]{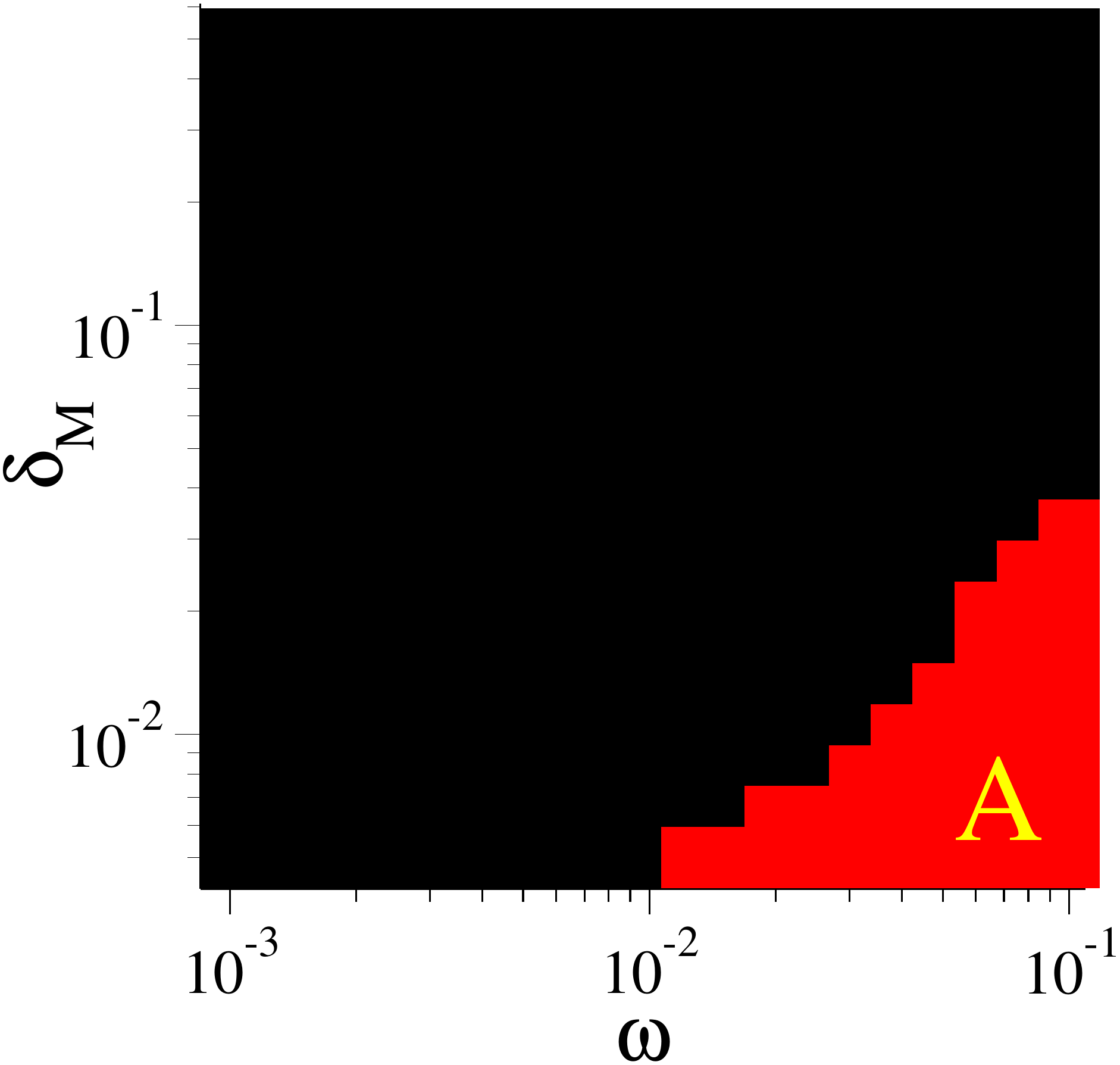}
\includegraphics[width=0.45\columnwidth]{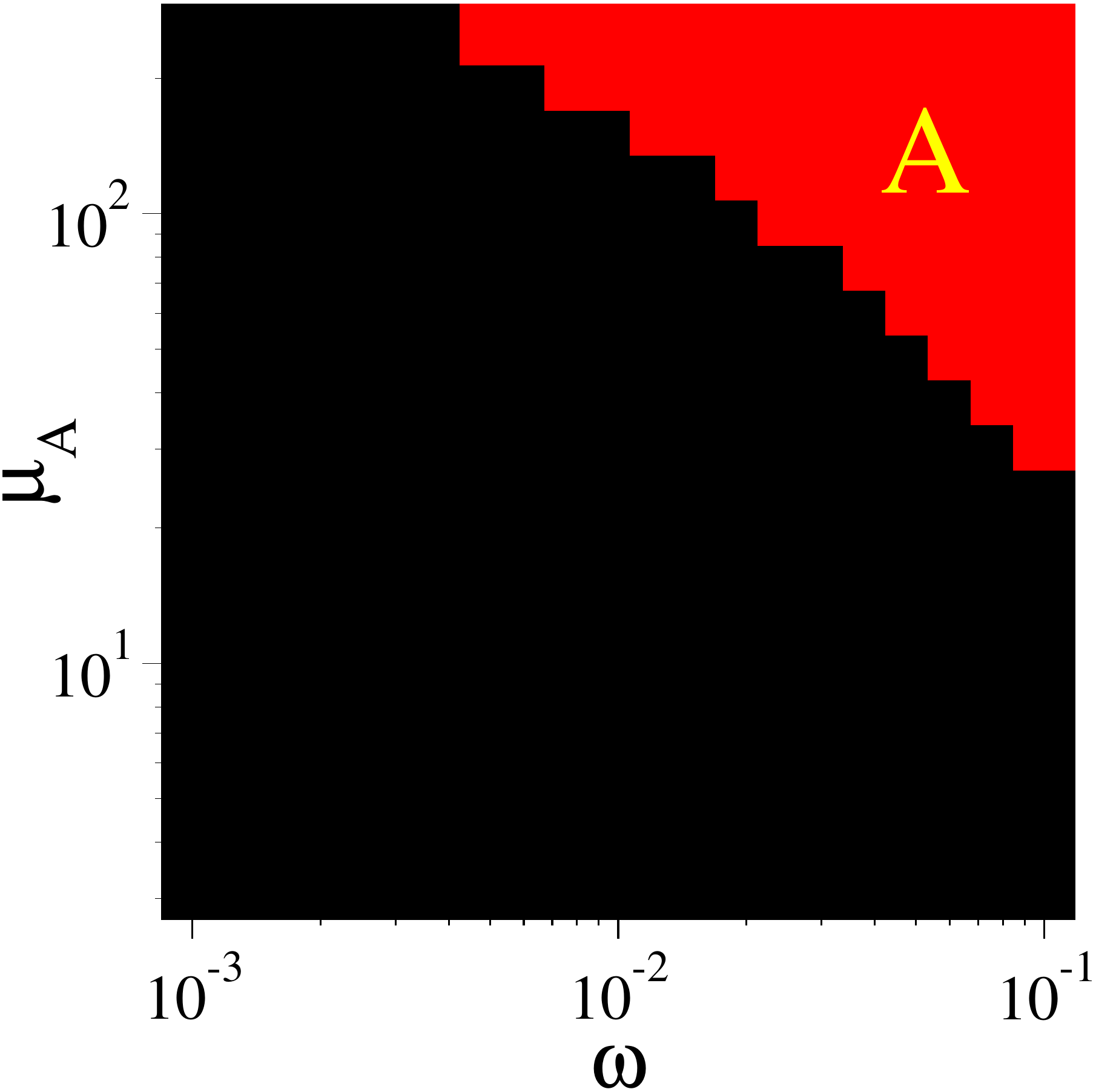}

\includegraphics[width=0.45\columnwidth]{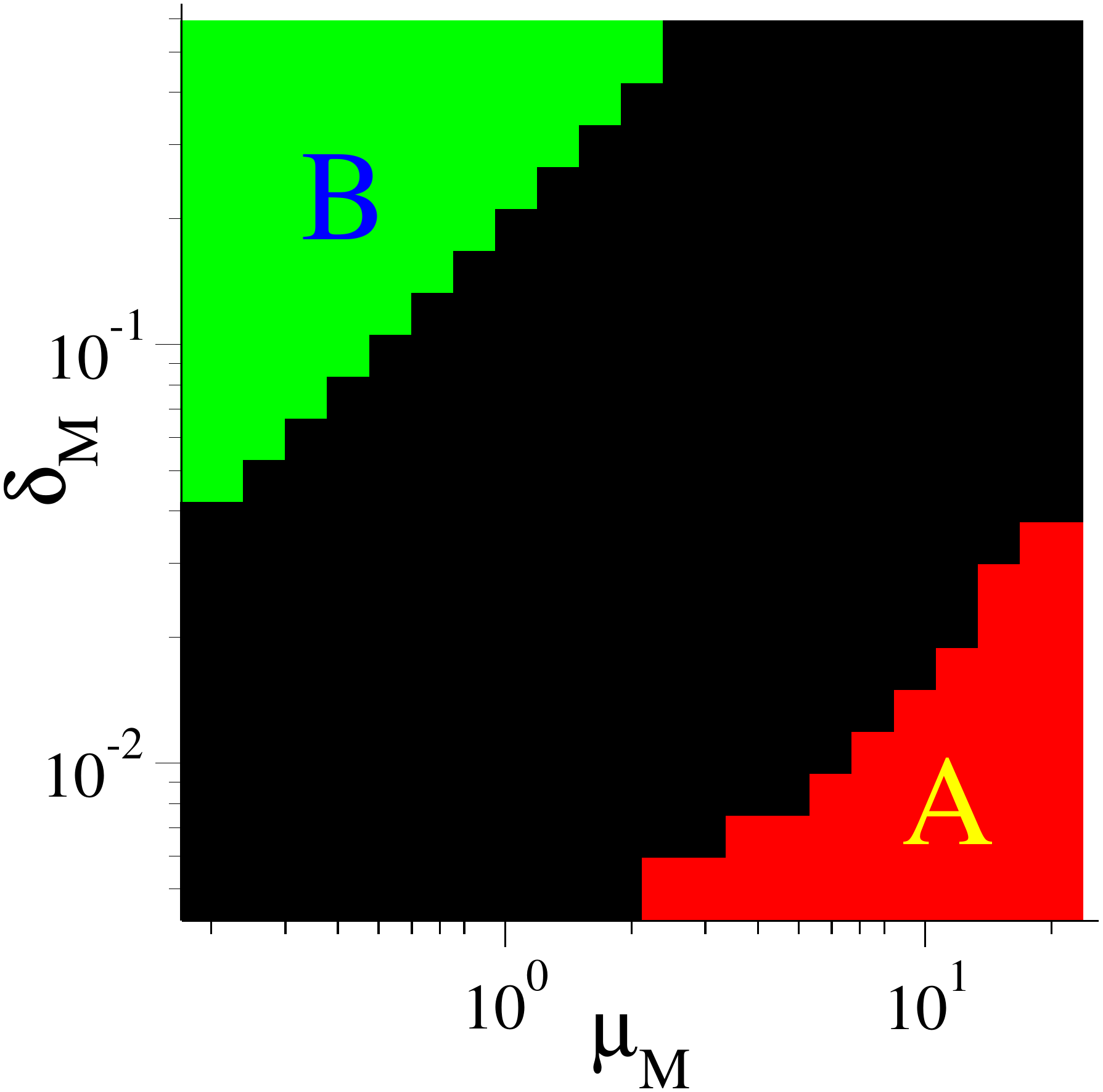}
\includegraphics[width=0.45\columnwidth]{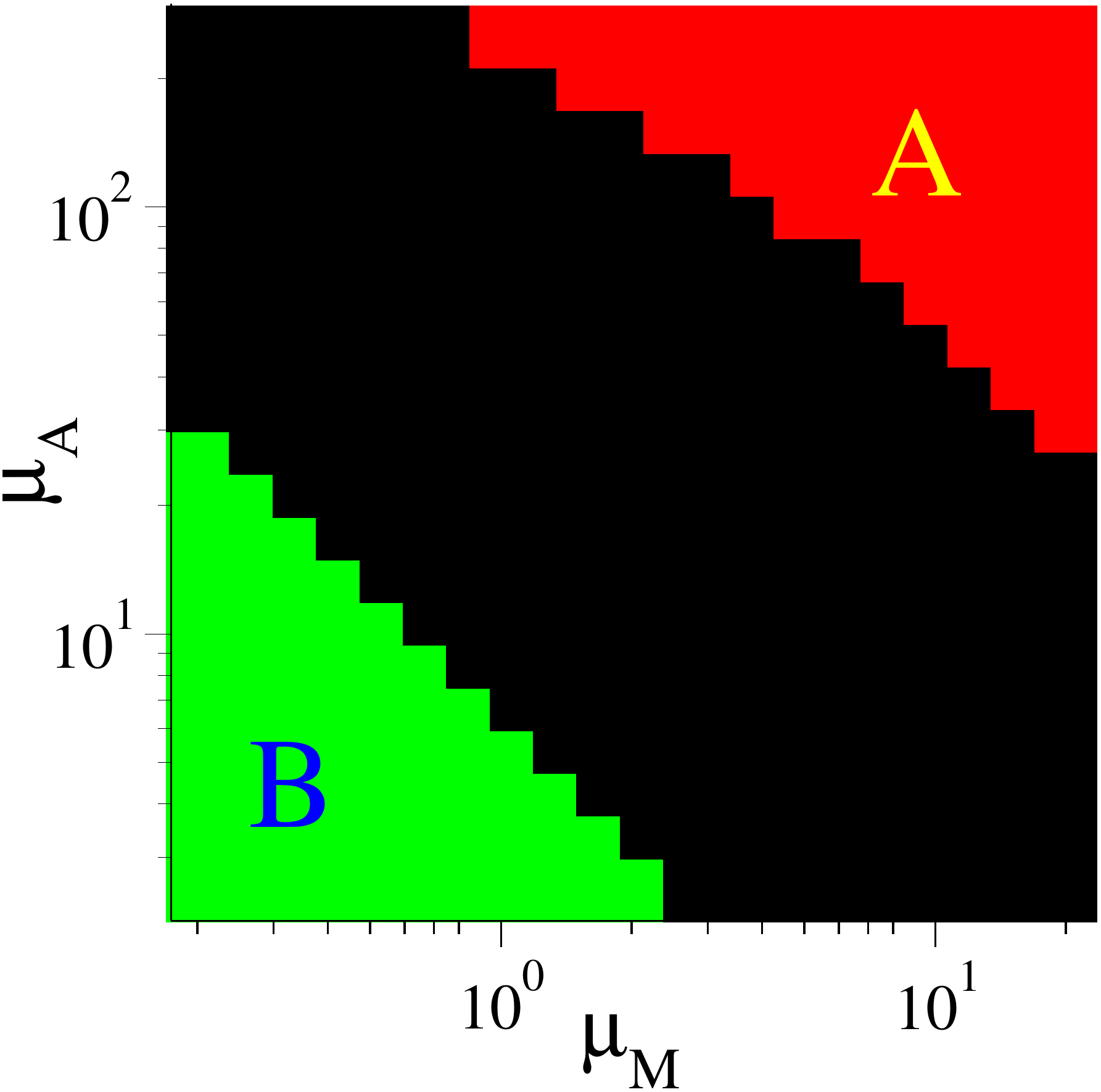}
\includegraphics[width=0.45\columnwidth]{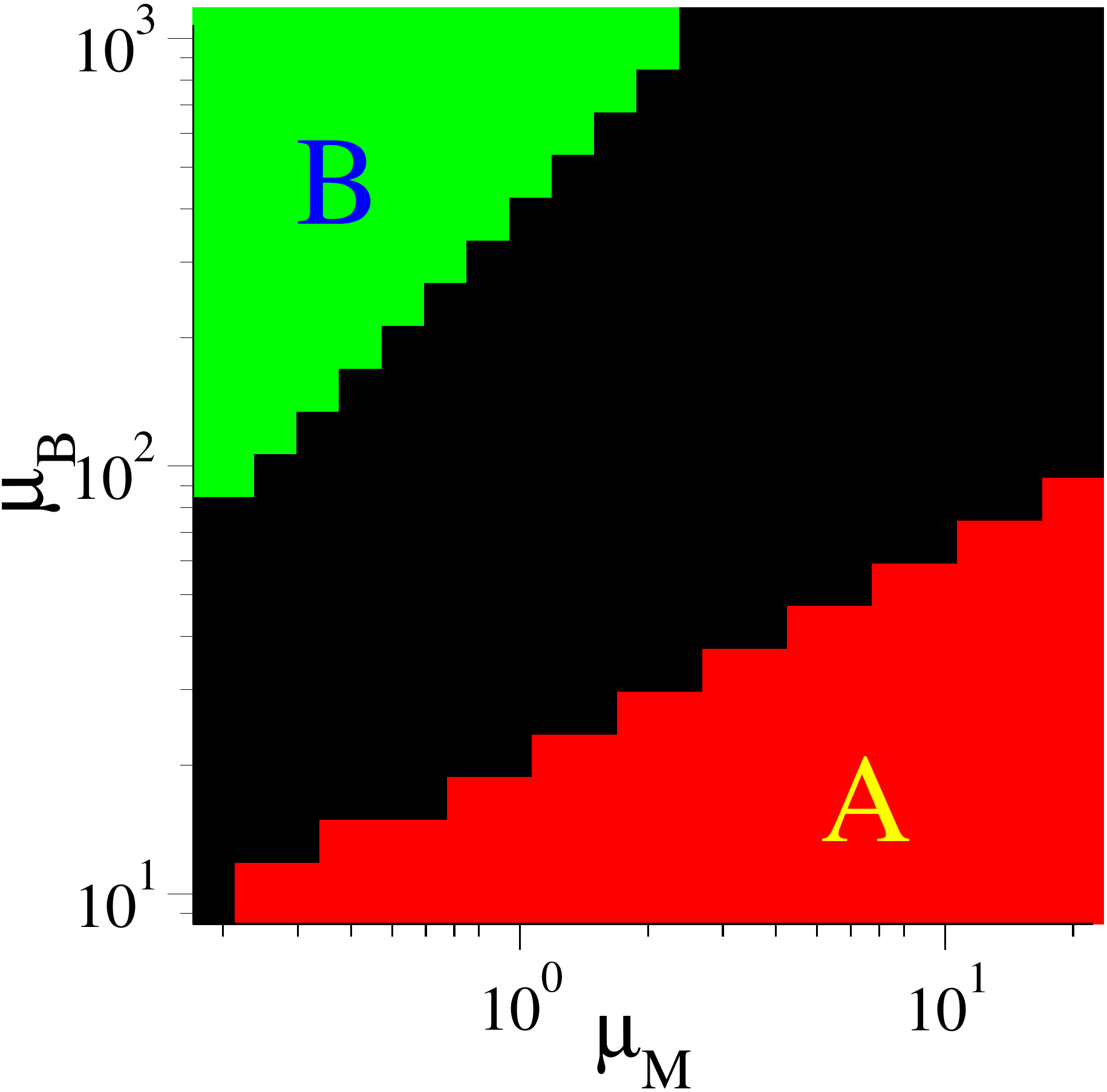}

\includegraphics[width=0.45\columnwidth]{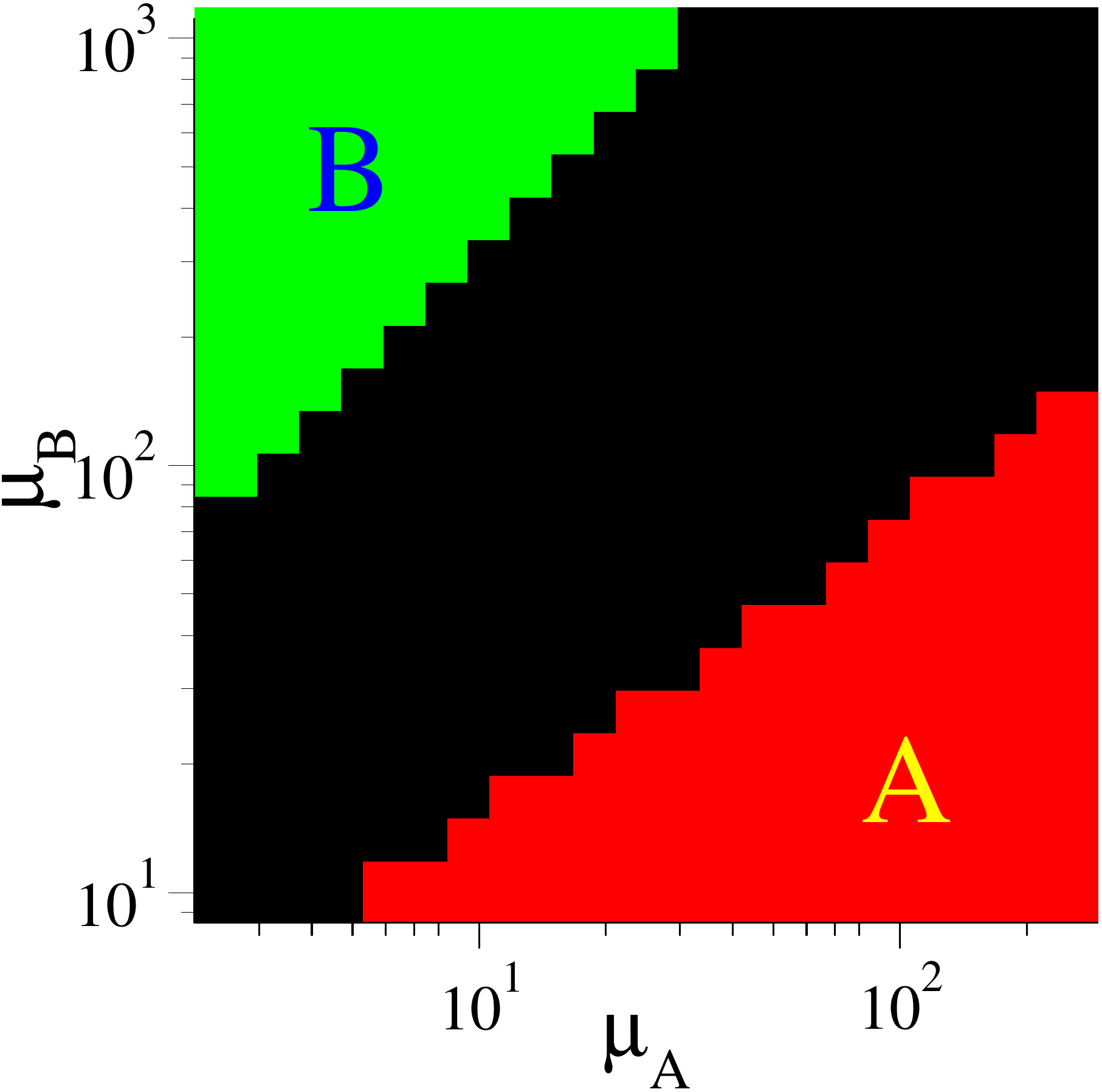}
\includegraphics[width=0.45\columnwidth]{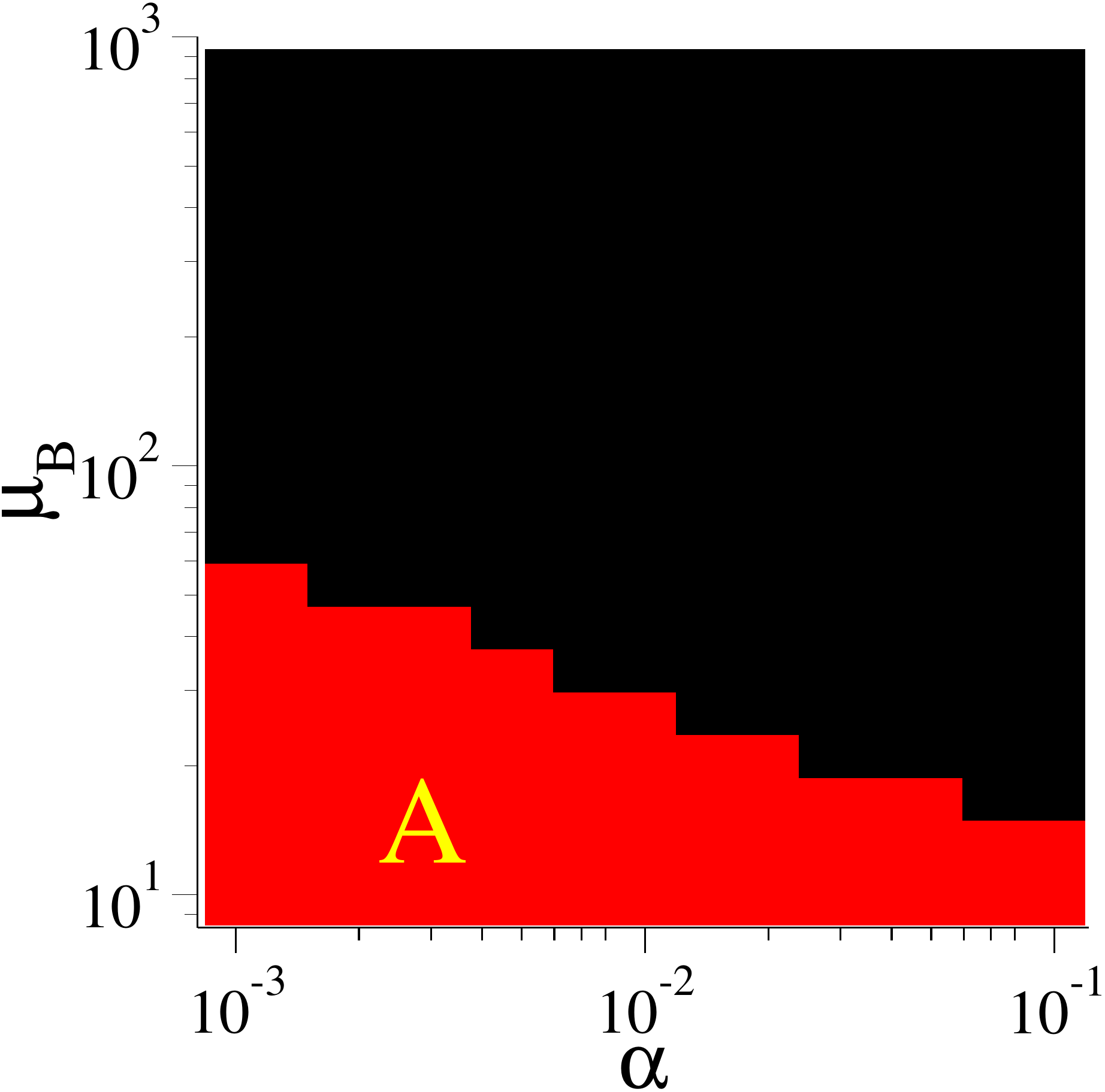}
\includegraphics[width=0.45\columnwidth]{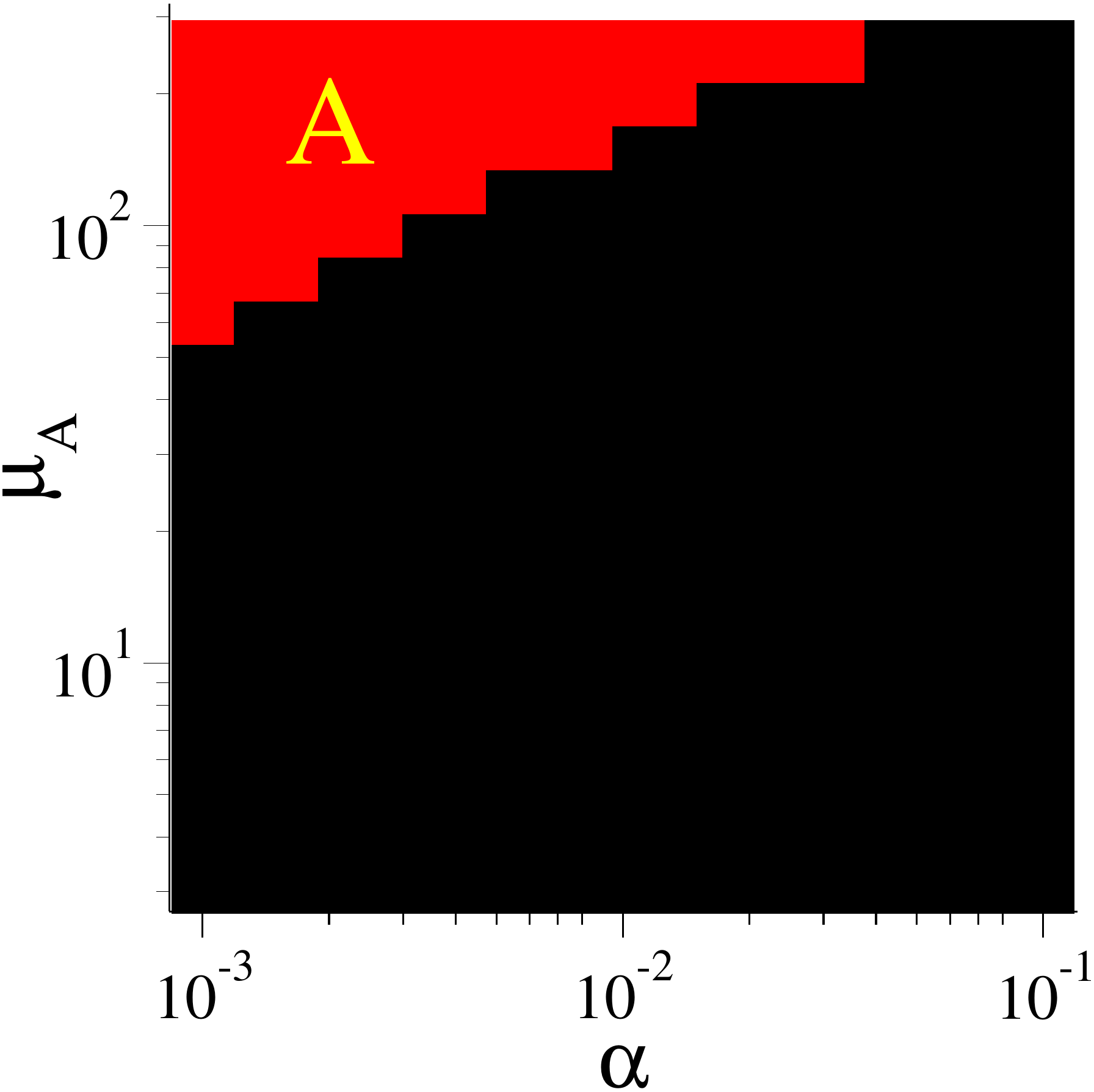}

\includegraphics[width=0.45\columnwidth]{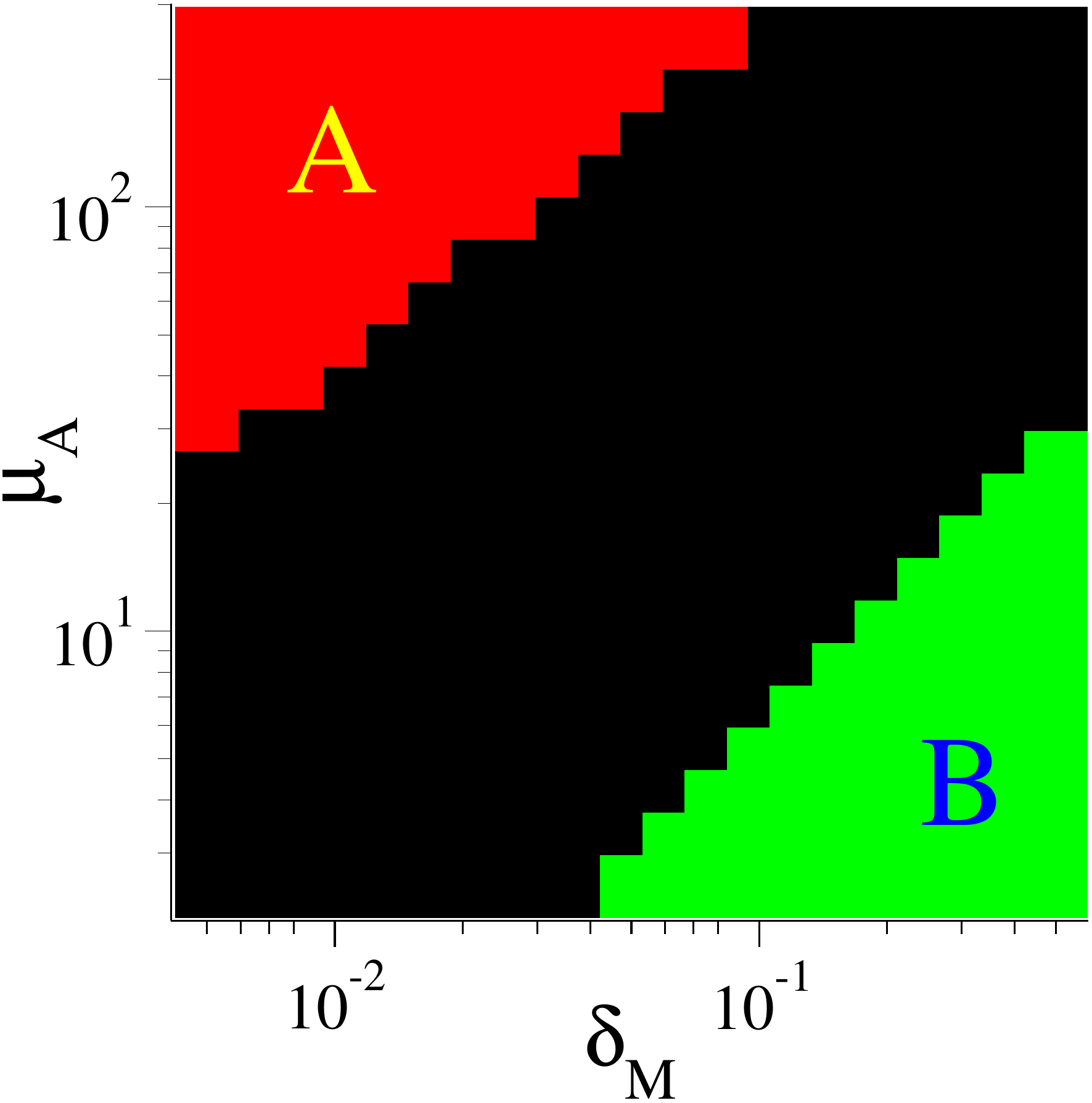}
\includegraphics[width=0.45\columnwidth]{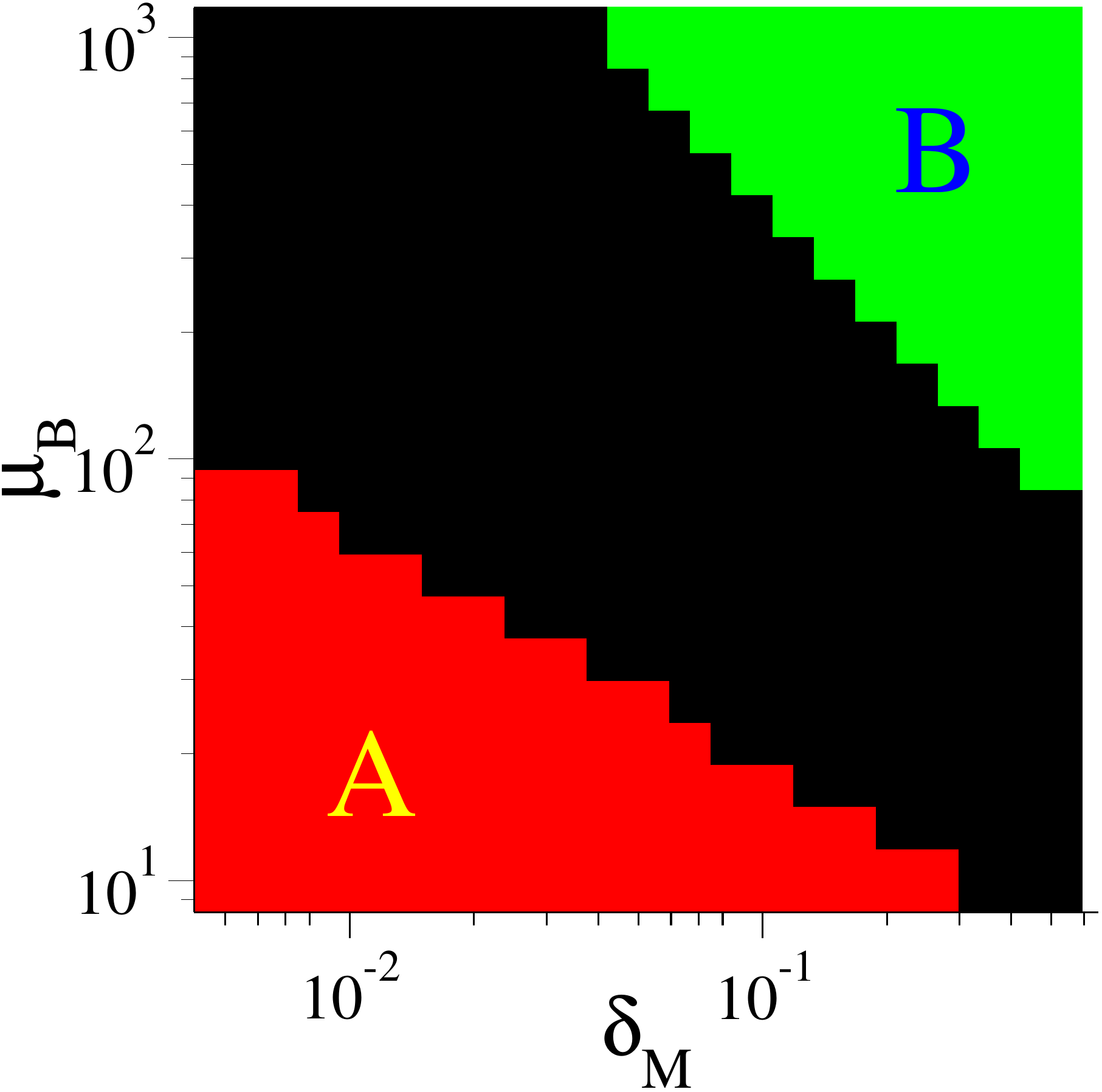}
\includegraphics[width=0.45\columnwidth]{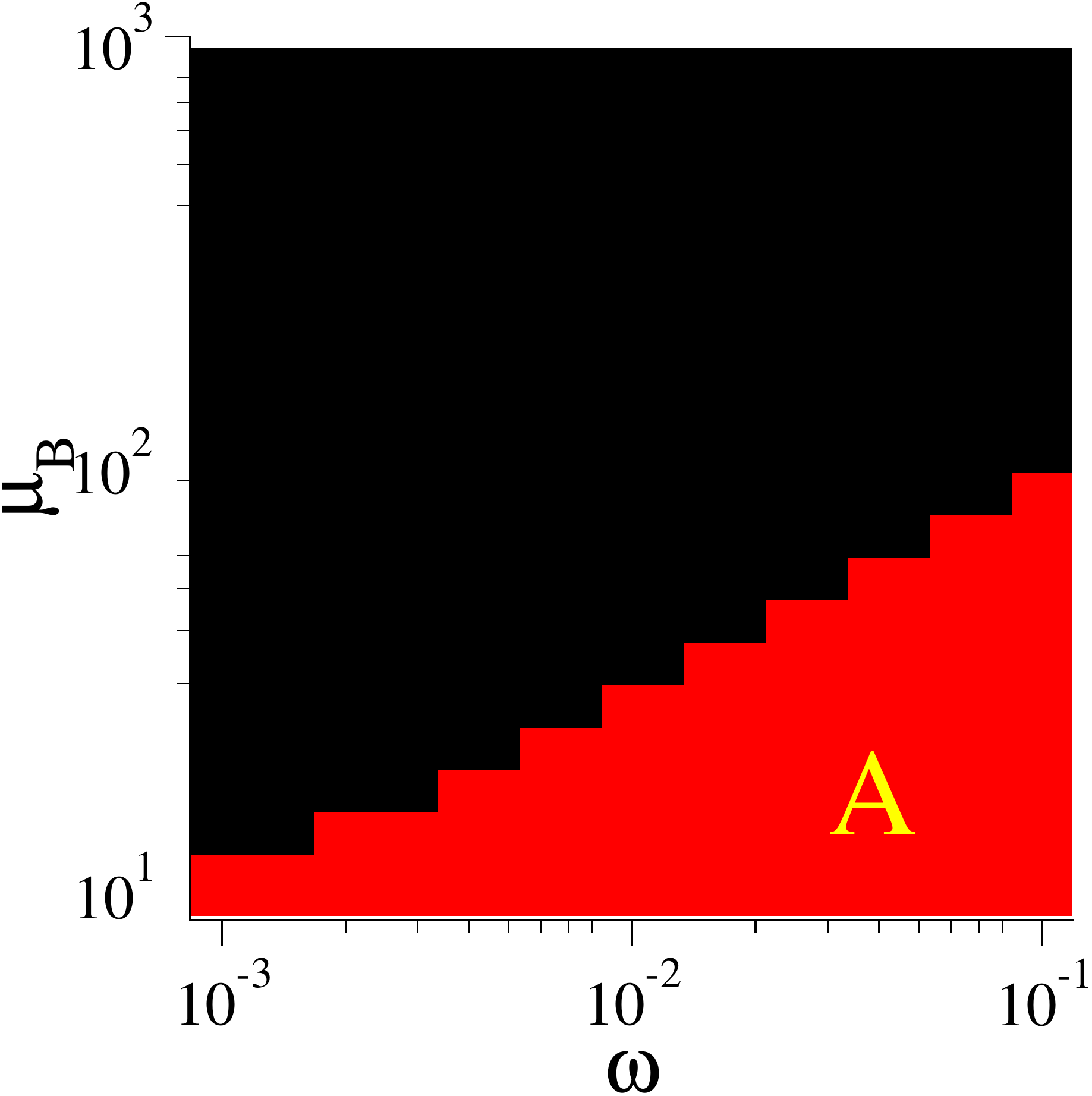}
\caption{Slices of phase diagrams of the model \eqref{suppl:full_model}
depicting various phases as a function of the six ``most relevant"
parameters $\alpha$, $\omega$, $\mu_M$, $\mu_A$, $\mu_B$, $\delta_A$
and $\delta_B$.  The black area corresponds to the pulsating domain,
while the red (resp. green) to steady states characterized by high
concentration of A (resp. B).}
\label{fig_phased}
\end{figure*}

Figure~\ref{fig_phased} shows a plot of ``phase diagrams" of the system.
To compute it we fixed the parameters to the following values $\mu_M^{-1}
= 0.5$, $\delta_M^{-1} =20$, $\mu_A^{-1} = 0.04$, $\mu_B^{-1} =
10^{-2}$, $\delta_A^{-1} =\delta_B^{-1} = 10^3$, $\delta_{A\! B}^{-1}
= 10$, $\gamma_{A\! B}^{-1}= 10$, $\omega^{-1} = 100$, $[A]_0 = 1$.
Six of these parameters $\alpha$, $\omega$, $\mu_M$, $\mu_A$,
$\mu_B$, $\delta_A$ and $\delta_B$ were varied two at a time while
keeping four of them fixed (recall that $\alpha = \omega/[A]_0$).
This procedure generates $6\cdot 5/2 =15$ two dimensional slices of
the phase diagram. The analysis consists in numerically integrating
Eqs.~\eqref{suppl:full_model} for every set of input rates while
identifying if the solution is pulsating or stationary.  The pulsating
domain is show as black in Fig.~\ref{fig_phased}. The axes in the phase
diagrams in Fig.~\ref{fig_phased} are in logarithmic scale and each
axes covers a variation of two orders of magnitude centered around the
values of rates given above. Hence the selected point is rather far from
the phase boundaries.  Apart from the pulsating solution we distinguish
two types of stationary solutions with high $A$ (high $B$) shown as red
(green) in Fig.~\ref{fig_phased}. In these phases one of the two proteins
has typically much higher concentration than the other.

%%%%%%%%%%%%%%%%%%%%%%%%%%%%%%%%%%%%%%%%%%%%%%%%%%%%%%%%%%%%%%%%%%%%%%%%%%%%
%%%%%%%%%%%%%%%%%%%%%%%%%%%%%%%%%%%%%%%%%%%%%%%%%%%%%%%%%%%%%%%%%%%%%%%%%%%%

\FloatBarrier

\clearpage

%%%%%%%%%%%%%%%%%%%%%%%%%%%%%%%%%%%%%%%%%%%%%%%%%%%%%%%%%%%%%%%%%%%%%%%%%%%%
\begin{figure}[t]
\includegraphics[width=1.0\columnwidth]{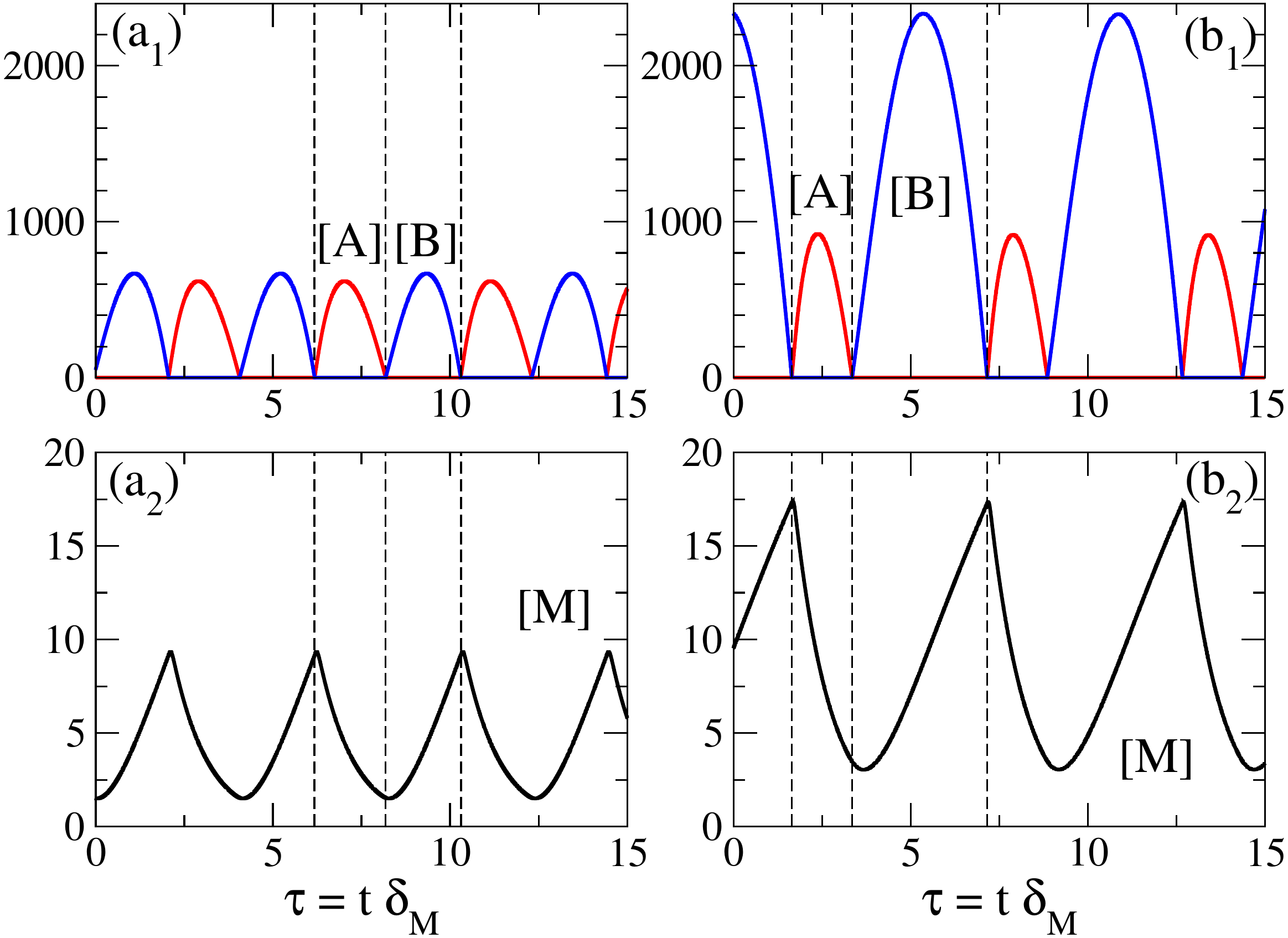}
\includegraphics[width=1.0\columnwidth]{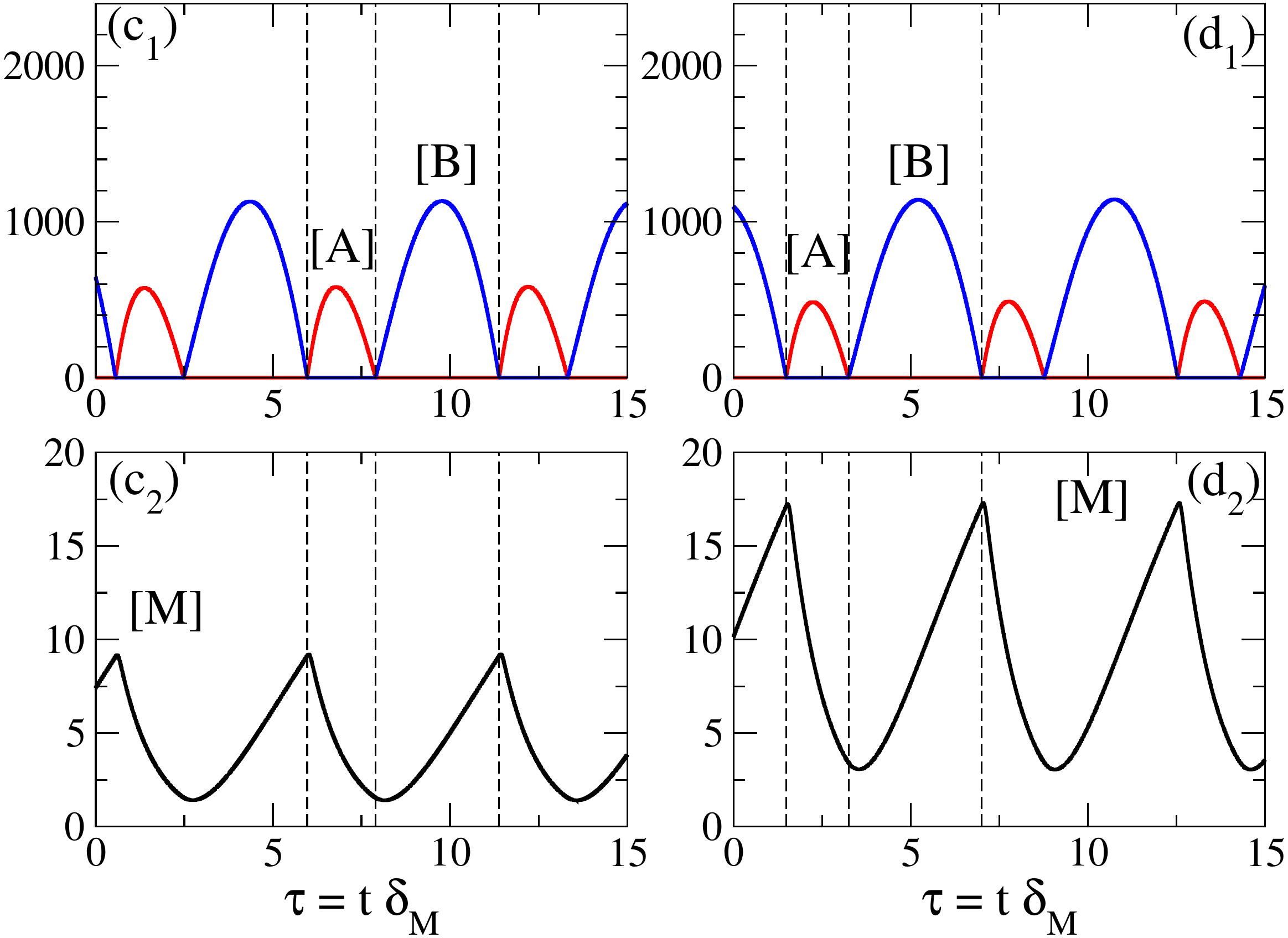}
\vspace{-6mm}
\caption{Plots of protein concentrations vs. time (a$_1$-d$_1$) and of
mRNA concentrations vs. time (a$_2$-d$_2$) for four different sets of rate
constants. In the case (a) the rates are those reported in Fig.~2 of the
main paper. (b,c and d) are obtained by (a) by a change of a single rate
constant as follows: (b) $\mu_B$ is doubled ($\mu_B^{-1}=0.00075$),
(c) $\omega$ is halved ($\omega^{-1}=200$) (d) $\mu_A$ is halved
($\mu_A^{-1}=0.1333$).}
\label{fig_param_variation}
\end{figure}
%%%%%%%%%%%%%%%%%%%%%%%%%%%%%%%%%%%%%%%%%%%%%%%%%%%%%%%%%%%%%%%%%%%%%%%%%%%%

Figure~\ref{fig_param_variation}(a-d) shows the effect on change of
rates on the concentrations of the proteins and mRNA in the pulsating
regime. Fig.~\ref{fig_param_variation}(a) reproduces the same rates as
in Fig.~2 of the main paper, while the cases (b), (c) and (d) correspond
to variations of a single rate with respect to the case (a). The duration
of the $A$ peaks is quite robust against the parameters variation, while
$T_b$ varies: (a) $T_a=2.01$, $T_b=2.13$ (b) $T_a=1.71$, $T_b=3.81$
(c) $T_a=1.93$, $T_b=3.48$ and (d) $T_a=1.76$, $T_b=3.76$.  In the case
(b) the $B$ protein synthesis rate $\mu_B$ is doubled with respect of
(a). This has a strong effect in the height and duration of the peaks of
$B$, but a milder effect on the duration of the peaks of $A$. Halvening
the value of (c) $\omega$ and of (d) $\mu_A$ is also affecting strongly
the peaks of $B$. While the duration of the peaks of $A$ is robust, their
height is not. This appears to be mostly affected by a change in $\mu_B$,
in agreement with the analysis of the reduced model in the next section.

\section{Analysis of the reduced model}

We present here the details of the analytical calculations for the
durations of the A and B phases for the reduced model:
\begin{subnumcases}{}
\frac{dg}{d\tau}\,   = \Omega (1-g)       -\sigma g a \label{equG}\\
\frac{dm}{d\tau}     = \quad g     \qquad -       m   \label{equM}\\
\frac{da}{d\tau}\,   = \quad k_a  m \quad -       a b \label{equA}\\
\frac{db}{d\tau}\,   = \quad k_b   \qquad -       a b \label{equB}
\end{subnumcases}

As shown in Fig.~2 of the main text two phases can be identified in
the pulsating domain: in one phase, $[A]$ is peaked and $[B]$ is small,
while in the other phase $[B]$ is peaked and $[A]$ small. We refer to
these as to the A-phase and to the B-phase, respectively. The two phases
are due to the mutual sequestration of $A$ and $B$. In the A-phase the
$G_A$ gene is strongly repressed, the mRNA synthesis is stopped and the
mRNA concentration $m(t)$ decreases due to degradation. As long as $k_a
m > k_b$ (rescaled variables and parameters, see Eqs.~(\ref{equA}) and
(\ref{equB})) the production of $a$ dominates over the production of
$b$. Once the mRNA concentration drops and $k_a m < k_b$, the production
of $B$ becomes dominant and the concentration of $A$ starts decreasing
till $A$ is completely sequestered out of the system and the transition to
the B-phase is made. In the B-phase the protein $A$ is released from its
gene $G_A$ promoter site. The absence of repression produces a rise in
the mRNA concentration; when the mRNA concentration reaches the threshold
value $k_am = k_b$, the concentration of $B$ starts decreasing and one
is back to the A-phase again.

We compute now the duration of the two phases.  Let us start from the
A-phase.  A first assumption is that the gene is constantly repressed
when the concentration of $A$ is high, hence $g=0$. We can thus eliminate
the variable $g$ from Eq.~(\ref{equM}) to obtain the solution:
\begin{equation}
m(\tau) = m_A\, e^{-\tau}
\label{MinAphase}
\end{equation}
where we set the origin of time $\tau=0$ at the beginning of
the A-phase. Using the same notation as the paper $m_A$ and $m_B$ 
indicate the mRNA concentrations at the beginning of the A- and of
the B-phases. A second assumption is that in the A-phase the
concentration of $B$ is stationary hence $db/dt=0$, which implies $k_b=ab$
from Eq.~(\ref{equB}). Substituting this into Eq.~(\ref{equA}) and using
(\ref{MinAphase}) we get the following equation for the evolution of $a$:
\begin{equation}
\frac{da}{d\tau} =	k_a m_A\, e^{-\tau} - k_b
\label{dadt_approx}
\end{equation}
The solution of the previous equation with initial condition $a(0)=0$ is
\begin{equation}
a(\tau) = k_a m_A \, \left( 1 - e^{-\tau} \right)  - k_b \tau
\label{eq:a_tau}
\end{equation}
which is a function with a single maximum beyond which it decreases
monotonically and it becomes negative at long times, which is obviously
an unphysical result.  We can estimate the duration of the
A-phase from the requirement $a(T_a)=0$, which gives:
\begin{equation}
\frac{T_a}{ 1 - e^{-T_a}} = \frac{k_a m_A }{k_b} 
\label{eqTa}
\end{equation}
For the B-phase we assume that the concentration of free $A$ in
solution is very small so that the binding rate to the gene promoter
site is negligible. We can approximate Eq.~(\ref{equG}) with $dg/d\tau
\approx \Omega (1-g)$, from which we get the following solution:
\begin{equation}
g(\tau) = 1 - e^{-\Omega (\tau-T_a)} 
\label{g_phaseB}
\end{equation}
where we used the initial condition $g(T_a)=0$ in the B-phase, $T_a
\leq \tau \leq T_a + T_b$.  We approximate further the previous expression
to the first order in the exponential:
\begin{equation}
g(\tau) \approx \Omega (\tau-T_a) 
\label{g_phaseB2}
\end{equation}
and which is valid for $\tau-T_a \ll \Omega^{-1}$. We now plug  in the
previous expression into Eq.~(\ref{equM}) and solve it to get for the
mRNA concentration in the B-phase ($T_a \leq \tau \leq T_a+T_b$):
\begin{eqnarray}
m(\tau) &=& \Omega \left(\tau - T_a - 1\right) +
e^{-\left(\tau-T_a \right)} \left(m(T_a)+ \Omega\right) \nonumber\\
        &=& \Omega \left(\tau - T_a - 1\right) +
e^{-\left(\tau-T_a \right)} \left(m_A  e^{-T_a} + \Omega\right)
\label{MinBphase}
\end{eqnarray}
where we have used Eq.~(\ref{MinAphase}): $m(T_a) = m_A  e^{-T_a}$.
We proceed as done for the A-phase. We assume that $a$ is stationary
in the B-phase, i.e. $da/d\tau =0$ which yields $k_a m = ab$
(Eq.~(\ref{equA})). Substituting this result in Eq.~(\ref{equB}) we get
the following Equation for the growth of $b$:
\begin{equation}
\frac{db}{d\tau} = k_b - k_a m
\end{equation}
with $m(\tau)$ given by Eq.~(\ref{MinBphase}). 
Using the initial condition $b(T_a) =0$ we get:
\begin{eqnarray}
b(\tau) &=& k_b \left(\tau-T_a \right) - k_a 
\Omega \left[ \frac{1}{2} \left( \tau - T_a \right)^2 
- \left( \tau - T_a \right) \right] \nonumber\\
&-& k_a \left(1 - e^{-(\tau-T_a)}\right)
\left(m_A  e^{-T_a} + \Omega \right)
\end{eqnarray}
We obtain the length of the B-phase from the requirement 
that $b(T_a + T_b) =0$, which leads to the following relation
\begin{equation}
\left(k_b+k_a\Omega\right)T_b = \frac{1}{2} k_a\Omega T_b^2 +
k_a \left(1 - e^{-T_b}\right)
\left(m_A  e^{-T_a} + \Omega \right)
\label{BinBphase}
\end{equation}
An additional relation is obtained by requiring that at the end of
B-phase: $m(T_a+T_b)=m_A $ which yields from Eq.~(\ref{MinBphase}):
\begin{equation}
m(0) = \Omega \frac{ -1+ T_b + e^{-T_b}}{1 - e^{-T_a - T_b }}
\label{pbc}
\end{equation}
Inserting the previous equation in Eq.~\eqref{eqTa} we get:
\begin{equation}
\frac{T_a}{e^{T_a}-1} =  \beta \frac{-1+T_b+e^{-T_b}}{e^{T_a}-e^{-T_b}}
\label{per1_final}
\end{equation}
where we defined $\beta \equiv k_a \Omega/k_b$.
We now use Eq.~\eqref{eqTa} to get an expression for $m(0)$ which we substitute
in \eqref{BinBphase} to get:
\begin{equation}
\frac{T_a}{e^{T_a}-1} = 
\frac{\beta \left(T_b - \frac{T_b^2}{2}\right) +  T_b}{1 - e^{-T_b}} - \beta
\label{per2_final}
\end{equation}
We also note that for $\beta=2$ the exact solution of
Eqs.~(\ref{per1_final}) and (\ref{per2_final}) is $T_a=T_b=2$, i.e. the
two phases have equal duration.  For $\beta > 2$ ($\beta < 2$) one has
$T_a > T_b$ ($T_a < T_b$). In terms of the original kinetic constants,
the parameter $\beta$ reads:
\begin{equation}
\beta = \frac{\omega \mu_M \mu_A }{\mu_B \delta_M^{\, 2}}
\end{equation}
It characterizes the relative importances of the A-phase and B-phase.
The A-phase dominates if $\omega$ (unbinding rate of the repressor A
from its gene), $\mu_M$ (mRNA synthesis rate) or $\mu_A$ (protein A
synthesis rate) are large. The B-phase is favored when $\mu_B$ (protein
B synthesis rate) or $\delta_M$ (mRNA degradation rate) are large.

\subsection{On the robustness of $T_a$}

We analyze now the dependence of $T_a$ and $T_b$ on $\beta$.
Eq.~\eqref{per2_final} is of the form
\begin{equation}
f_1(T_a) = f_2(T_b,\beta)
\label{func_eq}
\end{equation}
where $f_1$ and $f_2$ are the following functions:
\begin{equation}
f_1(x) = \frac{x}{e^x-1}
\end{equation}
and
\begin{equation}
f_2(x,\beta) = \frac{\beta \left(x - \frac{x^2}{2}\right) +  x}{1 - e^{-x}} - \beta
\end{equation}
For any $x > 0$, the function $f_1(x)$ satisfies $0 < f_1(x) < 1$.  This
implies that $T_b$, solution of \eqref{per1_final} and \eqref{per2_final}
must be such that $0<  f_2(T_b,\beta) < 1$.

%%%%%%%%%%%%%%%%%%%%%%%%%%%%%%%%%%%%%%%%%%%%%%%%%%%%%%%%%%%%%%%%%%%%%%%%%%%%
\begin{figure}[t]
\includegraphics[width=0.9\columnwidth]{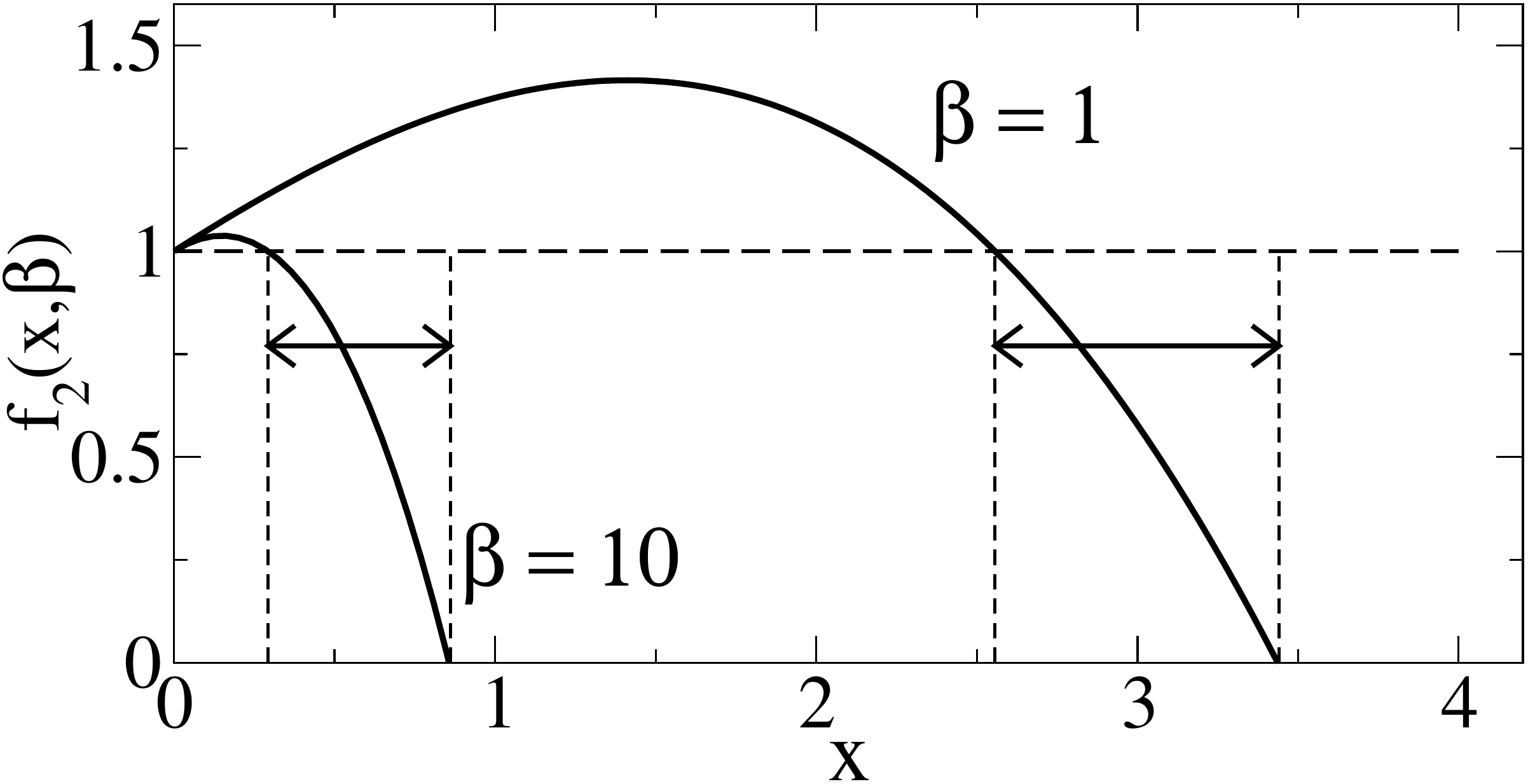}
\caption{Plot of the function $f_2(x,\beta)$ showing that 
the solutions for $T_b$ of Eq.~\eqref{func_eq} are in a limited
range of $x$.}
\label{fig_last}
\end{figure}
%%%%%%%%%%%%%%%%%%%%%%%%%%%%%%%%%%%%%%%%%%%%%%%%%%%%%%%%%%%%%%%%%%%%%%%%%%%%

Figure~\ref{fig_last} shows a plot of $f_2(x,\beta)$ for two values of
$\beta$. For large values of $x$, the function becomes negative and
its value satisfies $0< f_2(T_b,\beta) < 1$ only for a limited
range of $x$. This range varies strongly with $\beta$, which implies a
variation of $T_b$ with $\beta$. The analysis of $f_2(x,\beta)$
shows that $\lim_{\beta \to 0} T_b = \infty$ and 
$\lim_{\beta \to \infty} T_b =0$.

To proceed further we combine \eqref{per1_final} and \eqref{per2_final}
to eliminate $\beta$. We obtain:
\begin{equation}
\frac{T_a}{e^{T_a}-1} = 
\frac{T_b}{1-e^{-T_b} - 
\frac{T_b-\frac{T_b^2}{2}-1+e^{-T_b}}{-1+T_b+e^{-T_b}} \left(e^{T_a}-e^{-T_b}\right)}
\label{Ta_vs_Tb}
\end{equation}
In the limit $T_b \to 0$ (large $\beta$) the previous relation becomes: 
\begin{equation}
\frac{T_a}{e^{T_a}-1} = \frac{3}{e^{T_a}+2}
\label{eqTa_beta-infty}
\end{equation}
which has as unique solution $T_a \approx 2.149$.
In the opposite limit $T_b \to \infty$ we get from \eqref{Ta_vs_Tb}:
\begin{equation}
\frac{T_a}{e^{T_a}-1} = 2 \, e^{-T_a}
\label{eqTa_beta0}
\end{equation}
which has as solution $T_a \approx 1.594$. Hence this analysis shows
that while $T_b$ is unbounded and assumes any positive values when
$\beta$ is varied, $T_a$ is bounded in the interval $[1.594,2.149]$.
As discussed in the paper, the changes in rate constants which could
potentially affect $T_a$ are compensated by a change in $m_A$, the mRNA
concentration at the beginning of the A-phase, such that the ratio
\begin{equation}
c \equiv \frac{k_a m_A}{k_b}
\label{eq:def_c}
\end{equation}
remains constant. Using Eq.~\eqref{eqTa}, we can compute a range for $c$
using the estimated range of values of $T_a$. The result is $2 \leq c
\leq 2.4$.

\subsection{On the amplitude of $A$}

One can get some insights on the amplitude of $A$ from the analysis 
of the simplified model. The maximum of $a$ is obtained from 
Eq.~\eqref{eq:a_tau}:
\begin{equation}
\max_\tau {a} = k_b \left( \frac{k_a m_A}{k_b} - 1 - \log \frac{k_a m_A}{k_b}
\right)
\end{equation}
which shows that this quantity is not robust. Indeed, we have shown that
the solution of Eqs.~\eqref{per1_final} and \eqref{per2_final} are
such that the ratio \eqref{eq:def_c} is robust. The maximum of $a$
depends on this ratio, but it is also is proportional to $k_b$.
Transforming back to the original concentration units we find for 
the peak of $A$
\begin{equation}
\max_t {[A]} = \frac{\mu_B}{\delta_M} \left( c - 1 - \log c \right)
\end{equation}
where $c$ is defined in \eqref{eq:def_c}. This suggests that,
besides fixing $\delta_M$ which determines the overall timescale,
to control the height of the peaks of $[A]$ one needs to control
$\mu_B$, the $B$ production rate.  This is consistent with the plots of
Fig.~\ref{fig_param_variation}: the height of the $A$ peaks is mostly
affected by a change in $\mu_B$ (case (b)).

%%%%%%%%%%%%%%%%%%%%%%%%%%%%%%%%%%%%%%%%%%%%%%%%%%%%%%%%%%%%%%%%%%%%%%%%%%%%
\begin{figure}[t]
\includegraphics[width=0.95\columnwidth]{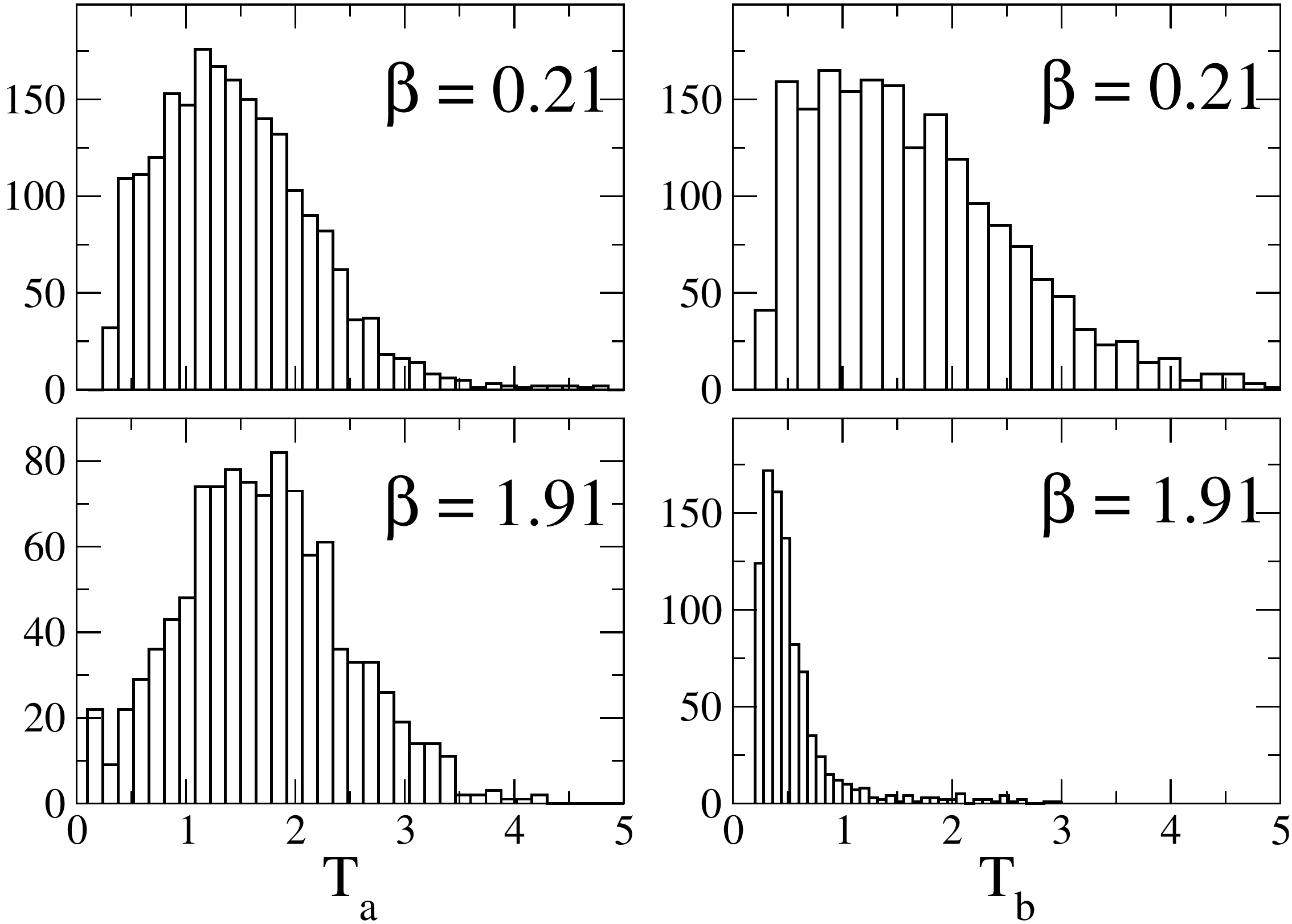}
\caption{Histograms of $T_a$ and $T_b$, the duration of the two
phases obtained from stochastic simulations with the Gillespie
algorithm~\cite{gill77} using two different choices of rates corresponding
to two values of $\beta$. While $\beta$ varies of almost an order of
magnitude the distibution of $T_a$ is weakly modified, whereas $T_b$
is strongly affected.}
\label{fig:stoch_distribution}
\end{figure}
%%%%%%%%%%%%%%%%%%%%%%%%%%%%%%%%%%%%%%%%%%%%%%%%%%%%%%%%%%%%%%%%%%%%%%%%%%%%

\section{Stochastic analysis}

We extended the analysis of the HAL to the stochastic regime, performing
simulations using the Gillespie algorithm. Typical outputs of these
simulations are given in Fig.~5 of the main text, which shows that the
protein concentrations evolve through peaks of variable duration and
height due to stochastic fluctuations. To quantify the variability
in the dimensionless durations of the two phases, we studied their
probability distribution for two different parameter sets, as shown in
Figure~\ref{fig:stoch_distribution}. The two top graphs are obtained
using the parameter values $\mu_M^{-1} = 1.11$, $\delta_M^{-1} =
16.67$, $\mu_A^{-1} = 0.59$, $\mu_B^{-1} = 0.05$, $\delta_A^{-1} =
\delta_B^{-1} = 10^3$, $\delta_{A\!B}^{-1} = 10$, $\gamma_{A\! B}^{-1}
= 0.02$, $\omega^{-1} = 100$, $[A]_0 = 1$ (and $\lambda_{A\! B}$ and
$\mu_M^A$ fixed as in Table I). These constants correspond to a value of
$\beta=0.21$. The two bottom graphs are generated using the same rates
except for $\delta_M^{-1} =50$, which corresponds to $\beta=1.91$. We
note that $\beta$ strongly influences the duration of the peaks of $B$;
conversely the distribution of $T_a$ is only weakly affected while $\beta$
is varied of almost an order of magnitude. This is consistent with the
deterministic analysis developed in the main text.

%  \begin{table}[htpb]
\begin{table}[t]
\centering
\begin{tabular}[c]{c|c|ccc}
      $\beta$ &Quantity & Stochastic & Deterministic & Analytical\\ \hline
      0.21 & $\langle T_a \rangle$ & 1.47 & 1.30 & 1.70 \\
           & $c_v(T_a)$ & 0.49 \\
           &$\langle T_b \rangle$ & 1.68 & 1.48& 10.00\\
                      & $c_v(T_b)$ & 0.60 \\ \hline
                      1.91 & $\langle T_a \rangle$ & 1.71& 1.44& 1.98\\
           & $c_v(T_a)$ & 0.44 \\                      
           &$\langle T_b \rangle$ & 0.51& 0.35& 0.50\\
           & $c_v(T_b)$ & 0.83 \\ \hline
    \end{tabular}
\caption{Average durations obtained from (i) stochastic simulations
of the chemical reactions in Fig.~\ref{BiochemicalReactions},
(ii) deterministic simulation of Eqs.~\eqref{suppl:full_model} and
(iii) Eqs.~\eqref{per1_final} and \eqref{per2_final} (analytical
approximation).}
\label{tab:indicat}
\end{table}

\FloatBarrier

The average values $\langle T_s \rangle$ and coefficients of
variation $c_v(T_s) = \sigma(T_s)/\langle T_s \rangle$, where
$\sigma(T_s)$ is the standard deviation of $T_s$, are given in
Table~\ref{tab:indicat}, which compares them to the values of $T_a$
and $T_b$ obtained from the deterministic simulations of the full
model [Eqs.~\eqref{suppl:full_model}], as well as from the analytical
approximation [Eqs.~\eqref{per1_final} and \eqref{per2_final}].
The values obtained confirm that $T_a$ is much less sensitive than
$T_b$ to $\beta$. For each value of $\beta$ and each average duration,
the agreement between the three estimates is reasonable except for
$\langle T_b \rangle$ at low $\beta$. This seems to indicates that
for some parameter sets with a low value of $\beta$, the analytical
approximation severerely overestimates $T_b$, perhaps because it
misses an ingredient leading to a faster dynamics.  However, this does
not affect our conclusion that $T_a$ is relatively constant, nor that
$T_b$ is largely tunable.  Table~\ref{tab:indicat} also shows that the
stochastic variability affects more the duration of the $B$ phase than
that of the $A$ phase.

Summarizing, the stochastic analysis of the system supports the
conclusions drawn from the study of the deterministic model:
while the A-phase appears to be rather robust, the duration of the 
B-phase is tunable and more subject to stochastic fluctuations.
One interesting issue to be left for future investigations is 
whether it is possible to find simple extensions of this genetic
module for which the stochastic fluctuations in $T_a$ can be 
further reduced.

\bibliography{my_bib}

\end{document}